\documentclass[final,3p,times,authoryear]{elsarticle}

\usepackage{amssymb, amsmath, amsthm}
\usepackage{url}
\usepackage{graphicx} 
\usepackage{caption}
\usepackage{subcaption} 
\usepackage{bm}
\usepackage{placeins}
\usepackage{here}
\usepackage{booktabs}
\usepackage{algorithm}
\usepackage{algpseudocode}
\usepackage{multirow}
\usepackage{xcolor}
\usepackage[bookmarks=true,bookmarksopen=true,bookmarksnumbered=true,bookmarksdepth=3,hidelinks]{hyperref}

\begin{document}

\begin{frontmatter}

\title{{3D Shape Recovery of Human-Made Space Objects from Light Curves with Image Denoisers as Priors}}

\author[1]{Daigo Kobayashi\corref{cor1}}
\ead{dkobayashi@ua.edu}

\author[2]{Carolin Frueh}
\ead{cfrueh@purdue.edu}

\cortext[cor1]{Corresponding author.}

\affiliation[1]{
    organization={Department of Aerospace Engineering and Mechanics,
                  The University of Alabama},
    city={Tuscaloosa},
    postcode={35487},
    state={AL},
    country={United States}
}

\affiliation[2]{
    organization={School of Aeronautics and Astronautics,
                  Purdue University},
    city={West Lafayette},
    postcode={47907},
    state={IN},
    country={United States}
}

\begin{abstract}
This study developed a {denoiser-regularized} framework for recovering the three-dimensional convex shape of {human-made space objects} from unresolved light curves. Such information supports object characterization for space situational awareness and mission support. However, conventional extended Gaussian image (EGI) recovery by nonnegative least squares and directional refinement can discard surface-normal information needed for curved geometries. The proposed framework instead estimates the EGI as an area image on an equirectangular grid, recovers it in a randomly rotated frame using a plug-and-play alternating direction method of multipliers with a domain-adapted DRUNet denoiser, and converts it into a convex polytope via scale-invariant support optimization with analytical gradients and volume normalization. Across seven faceted and curved shapes and 10 runs per condition, mean convex-hull intersection-over-union (IoU) at $30\,\mathrm{dB}$ rises from $0.615$ for the baseline to $0.872$, while mean Hausdorff and Chamfer distances fall from $0.198$ to $0.050$ and from $0.167$ to $0.041$, improving all three metrics on six of the seven shapes. Accuracy is largely retained under noise, with mean IoU decreasing only from $0.881$ to $0.854$ between noiseless measurements and a signal-to-noise ratio of $20\,\mathrm{dB}$. 
\end{abstract}

\begin{keyword}
Light-curve inversion
\sep Resident space objects
\sep Space situational awareness
\sep Extended Gaussian image
{
\sep Plug-and-play priors
\sep Alternating direction method of multipliers
}
\end{keyword}

\end{frontmatter}

\section{Introduction}
Accurate prediction of the orbital and attitude evolution of {human-made space objects} depends on the fidelity of the dynamical models and the information available to constrain them. Modeling non-gravitational forces and torques, in particular, requires knowledge of an object's shape, attitude, and surface properties. Uncertainty in these quantities can propagate into trajectory predictions through coupled orbital and attitude dynamics \citep{fruh2014_coupled_orbit_attitude}. For cooperative spacecraft, design data and telemetry can provide this information. For uncooperative objects, however, these sources may be unavailable, making external observations necessary.

Optical photometry provides one source of information about objects that cannot be spatially resolved. Sampling an object's unresolved brightness over time yields a light curve, whose variations depend on its shape, attitude, and surface scattering. Light-curve inversion exploits this dependence to infer physical properties from photometric observations. Its development for asteroids includes general inversion theory \citep{kaasalainen1992a_general_theory_inversion} and methods for recovering shape, spin, and scattering parameters \citep{kaasalainen2001a_complete_inverse_problem}. These methods have also been adapted to artificial objects, extending photometric shape estimation beyond natural bodies \citep{bradley2014_lightcurve_inversion_shape}.

Despite these developments, the coupled contributions to brightness limit what light curves can uniquely determine. Different combinations of shape, attitude, and surface scattering can produce indistinguishable measurements \citep{kaasalainen2006_inverse_problems_generalized}. Observation geometry, defined by the relative positions of the Sun, object, and observer, also affects shape-feature observability \citep{friedman2022_observability_light_curve}. This geometry and the object's attitude determine the illumination and viewing directions in the body-fixed frame. Methods address these dependencies through different constraints on the unknown quantities. Filter-based approaches estimate shape, attitude, and trajectory jointly within prescribed shape parameterizations or candidate sets \citep{linares2012_inactive_space_object,linares2014_space_object_shape}. Learning-based methods constrain their outputs using training data, although much of the cited artificial-object work focuses on classification \citep{furfaro2018_space_objects_classification,allworth2021_transfer_learning_approach}. Beyond classification, \citet{tang2025_asteroid_shape_inversion} reconstruct asteroid shapes as convex-hull point clouds and identify non-convex regions from photometric discrepancies. These approaches differ in their estimation strategies, but each relies on physical information or model restrictions to constrain recovery.

Non-convexity introduces an additional physical difficulty through self-shadowing and mutual occlusion. A surface element may face the Sun and the observer yet be shadowed or obscured by another part of the object. Therefore, its brightness contribution depends on the spatial arrangement of the surface, not only on its area and normal direction. Accommodating non-convex geometry in a representation is consequently insufficient unless the forward model accounts for these illumination and visibility effects. For asteroid test bodies, \citet{kaasalainen2001b_shape_determination} showed that convex reconstructions can approximate the convex hulls of non-convex shapes. \citet{durech2003_photometric_signatures_highly} further demonstrated that the detectability of non-convex features depends strongly on asteroid morphology and solar phase angle. These findings concern the shapes and observing conditions investigated. They do not establish that convex models adequately reproduce artificial-object photometry under different component configurations and observing geometries.

These limitations motivate a clearly bounded shape-recovery formulation. Prescribing observation geometry, attitude, and a scattering model while reconstructing a convex shape is an established approach in artificial-object light-curve inversion \citep{vallverducabrera2021_inversion_shape_space,friedman2022_observability_light_curve}. Following this approach, this paper assumes known observation geometry, attitude history, and bidirectional reflectance distribution function (BRDF), including its reflection parameters. The recovered shape is constrained to be convex. Recovery of non-convex shapes and resolution of general non-uniqueness remain outside its scope. Within these constraints, this paper focuses on recovering mixed curved and faceted geometry from noisy light curves.

Even within the convex class, artificial-object geometry places distinct demands on the shape representation. Smooth-function expansions can describe global asteroid shapes compactly, but large planar regions and sharp features may require many expansion terms \citep{kaasalainen2001b_shape_determination}. Artificial objects can contain both faceted and curved components, as illustrated by the CubeSat, box-wing, and upper-stage models considered by \citet{bradley2014_lightcurve_inversion_shape}. Thus, their reconstruction requires a representation that accommodates planar facets, sharp edges, and smoothly varying surfaces within the same object.

The extended Gaussian image (EGI) describes surface geometry by assigning surface area to outward unit-normal directions on the unit sphere \citep{horn1984_extended_gaussian_images}. For a smooth, strictly convex surface with positive Gaussian curvature, this measure has a continuous density. For a polytope, the EGI consists of discrete masses equal to the facet areas. Mixed geometry can therefore contain both distributed and concentrated area components. Continuous area-density formulations underpin asteroid inversion, including spherical-harmonic representations used in practical recovery \citep{kaasalainen1992a_general_theory_inversion,kaasalainen1992b_practical_aspects_inversion}. For polyhedral geometry, the EGI also provides a foundation for reconstructing the surface from its facet normals and areas. Positive facet areas whose area-weighted normals sum to zero determine a convex polytope uniquely up to translation, provided the normals span three dimensions \citep{minkowski1989_allgemeine_lehrs_atze}. \citet{little1983_iterative_method_reconstructing} developed reconstruction through facet support values, which specify supporting-plane offsets, and subsequently connected the formulation to mixed-volume theory \citep{little1985a_extended_gaussian_images}. This geometric foundation separates inversion into two steps: estimating the area measure from photometric data and reconstructing its associated convex shape.

A practical limitation of discrete EGI-based inversion is the approximation error introduced by finite sampling of surface-normal directions. In candidate-normal implementations, surface areas are assigned to a prescribed set of directions that does not necessarily coincide with the target's true facet normals. Such implementations have recovered idealized artificial objects, although accuracy varies across target configurations \citep{bradley2014_lightcurve_inversion_shape}. Subsequent normal resampling and merging have refined EGI estimation, alongside improvements to the numerical optimization of facet supports \citep{robinson2022_light_curve_inversion}. These advances do not, however, ensure an exact geometric representation. Exact representation of a polytope requires the candidate set to include its true facet normals and the associated areas to match its facet areas. Directional mismatch introduces discretization error \citep{fan2020b_light_curve_simulation}, while curved surfaces require approximation of their distributed normal measure.

Measurement noise introduces a separate difficulty by obscuring weakly constrained components of the estimated area distribution. Previous methods have addressed this difficulty through additional observations and regularization. \citet{fan2020a_direct_light_curve} used an initial light curve to generate shape hypotheses and a second light curve to select among them. Regularization instead constrains area estimation by introducing prior preferences, which can affect the recovered geometry. In the examples of \citet{vallverducabrera2021_inversion_shape_space}, Tikhonov regularization rounded reconstructed shapes, whereas Lasso favored sharper, more concentrated facet structure. A regularization choice that improved one target could distort another. The remaining challenge is therefore to stabilize recovery from noisy light curves while preserving both localized area concentrations and extended distributions.

Image reconstruction offers an established framework for combining physical measurement models with prior information about the unknown distribution. The alternating direction method of multipliers (ADMM) uses variable splitting to separate data fitting from regularization, with the resulting subproblems coupled through an agreement constraint \citep{boyd2011_distributed_optimization_statistical}. In a standard regularized formulation, the prior enters through the proximal operator of an explicit penalty. Plug-and-play ADMM (PnP-ADMM) replaces this operator with an image denoiser, allowing prior information to be incorporated without specifying an analytical regularization function \citep{venkatakrishnan2013_plug_play_priors}. The denoiser need not correspond to the proximal operator of an explicit penalty. This separation makes the prior replaceable without changing the physical observation model. For EGI recovery, organizing the area coefficients as an image therefore offers a way to encode concentrated and distributed normal-area patterns through a domain-adapted denoiser, rather than selecting a single sparsity or smoothness penalty. The modeling choice shifts from prescribing an explicit regularizer to selecting and adapting a denoiser for the area distributions of interest.

Building on this framework, this paper treats EGI estimation as an image reconstruction problem by mapping sampled normal directions onto an equirectangular grid, with pixel values representing the associated surface areas. This rasterization retains the finite-normal approximation while enabling PnP-ADMM recovery with a domain-adapted denoiser. The subsequent support-recovery stage retains Little's mixed-volume principle while using a scale-invariant objective, first-order parameter updates, and explicit volume normalization. This paper evaluates the two-stage method on faceted, curved, and mixed convex targets at prescribed measurement-noise levels. Surface-distance and volumetric-overlap measures assess whether the recovered area information translates into accurate three-dimensional geometry.

\section{Light-Curve Model}
\label{sec:lightcurve}

Optical observation of a space object by a ground-based sensor involves a sequence of physical processes. Incident sunlight is reflected from the object's surface, propagates to the observer, and is collected and focused by a telescope. The detector integrates the received radiation to produce an unresolved brightness measurement{ $y_j^{\text{raw}}$ at time step $j$}. 

The spatial variation in reflected brightness over the object's surface is modeled by representing its shape as a polygonal mesh consisting of $N$ triangular facets. Under the geometrical-optics approximation, the unresolved brightness is computed by summing the reflected-light contributions from the individual facets \citep{fan2020b_light_curve_simulation}. {The normalized brightness at time step $j$ is given by}
\begin{align}
    {{y}_j
    = \frac{r_{\mathrm{topo},j}^2}{I_0} y_j^{\text{raw}}} =
    \sum_{i=1}^N
    \mu_{ij}  A_i \ f_r(\mathbf{p}_i,\bm{\omega}_{\ell,j},\bm{\omega}_{o,j}) \ (\mathbf{n}_i\cdot\bm{\omega}_{\ell,j}) (\mathbf{n}_i\cdot\bm{\omega}_{o,j}).
    \label{eq:brightness}
\end{align}
Here, $A_i$ and $\mathbf{n}_i$ are the area and outward unit normal of facet $i$, respectively. The unit vectors $\bm{\omega}_{\ell,j}$ and $\bm{\omega}_{o,j}$ denote the object-to-Sun and object-to-observer directions expressed in the body-fixed frame. Throughout this paper, the subscript $j$ indexes light-curve samples and the subscript $i$ indexes facets or surface-normal directions. Surface reflection is described by the bidirectional reflectance distribution function (BRDF), $f_r(\cdot)$, evaluated using the facet-dependent surface parameters $\mathbf{p}_i$. The binary factor $\mu_{ij}$ is one when facet $i$ is illuminated and visible, with no occlusion or shadowing by other parts of the object; otherwise, it is zero. The incident solar irradiance is denoted by $I_0$, and $r_{\mathrm{topo},j}$ is the object-to-observer range. Thus, the factor $I_0/r_{\mathrm{topo},j}^{2}$ accounts for the incident illumination and inverse-square propagation loss. Atmospheric transmission and instrument response are not explicitly included in the present model. {Because the factor $I_0/r_{\mathrm{topo},j}^{2}$ is known from the observation geometry, it is removed by multiplying $y_j^{\text{raw}}$ by $r_{\mathrm{topo},j}^{2}/I_0$, which yields the normalized brightness $y_j$ in Eq.~\eqref{eq:brightness}.}

Surface reflection is represented using a normalized Blinn--Phong BRDF \citep{blinn1977_models_light_reflection,duvenhage2013_numerical_verification_bidirectional}, which includes both diffuse and specular components:
\begin{align}
    f_r(\mathbf{p}, \bm{\omega}_\ell, \bm{\omega}_o) =
    \frac{C_d}{\pi} + C_s\cdot \frac{s+2}{2\pi}\cdot\frac{(\mathbf{n}\cdot\bm{\omega}_h)^s}{4(\mathbf{n}\cdot\bm{\omega}_\ell)(\mathbf{n}\cdot\bm{\omega}_o)}, \label{eq:blinn-phong}
\end{align}
where
\begin{align}
    \bm{\omega}_h = \frac{\bm{\omega}_{\ell}+\bm{\omega}_{o}}{\|\bm{\omega}_{\ell}+\bm{\omega}_{o}\|}
\end{align}
is the halfway vector between the illumination and viewing directions.
The surface parameter set is $\mathbf{p}=\{\mathbf{n}, C_d, C_s, s\}$, where $C_d$ and $C_s$ are the diffuse and specular reflection coefficients, respectively, and $s$ is the specular exponent. Larger values of $s$ produce a narrower specular lobe, whereas smaller values produce a broader response. The reflection coefficients satisfy:
\begin{align}
    C_d \geq0, \quad C_s \geq 0, \quad C_d + C_s \leq 1.
\end{align}

{Collecting the normalized brightness at $m$ observation times yields the normalized light curve,}
\begin{align}
    {\mathbf{y}} = [{y}_1, \ \dots, \ {y}_m]^{{\mathsf{T}}}. \label{eq:LC_normalized} 
\end{align}
Unless otherwise stated, ${\mathbf{y}}$ is used as the measurement quantity in the subsequent shape-recovery formulation and numerical simulations.

Synthetic light curves are computed using a custom Python implementation of the facet-summation model{~\citep{robinson2023_light_curve_simulation}}. For the convex objects considered here, the implementation evaluates the Blinn--Phong response of each illuminated and visible facet and sums its area-weighted contribution directly to obtain unresolved brightness, without rendering an intermediate resolved image.

Measurement noise is introduced by drawing a vector of independent standard Gaussian samples and rescaling it to attain a prescribed light-curve signal-to-noise ratio (SNR). Let $\mathbf g\sim\mathcal N(\mathbf 0,\mathbf I_m)$. The noisy light curve is constructed as
\begin{equation}
    \mathbf y_{\mathrm{noisy}}=\mathbf y+\frac{\|\mathbf y\|_2}{10^{\mathrm{SNR}_{\mathrm{dB}}/20}}\frac{\mathbf g}{\|\mathbf g\|_2}.\label{eq:noisy_lightcurve}
\end{equation}
Here, $\mathrm{SNR}_{\mathrm{dB}}$ is the prescribed ratio of the summed squared clean-signal values to the summed squared noise values, expressed in decibels. The scaling ensures that this SNR is attained exactly for each realization.

\section{{Denoiser-Regularized Convex Shape Recovery}} \label{sec:proposed-method}

{Convex shape recovery from a light curve is traditionally performed in two stages: estimation of the extended Gaussian image (EGI), followed by reconstruction of the associated convex polytope. The proposed method retains these stages but estimates the EGI as a denoiser-regularized area image and converts it into a polytope by scale-invariant support optimization. Both stages use the normalized light curve $\mathbf{y}$ defined by Eq.~\eqref{eq:LC_normalized} and assume that the body-frame directions $\bm{\omega}_{\ell,j}$ and $\bm{\omega}_{o,j}$ and the reflectance parameters $C_d$, $C_s$, and $s$ are known, with the reflectance parameters spatially uniform.}

\subsection{EGI Formulation of Light-Curve Inversion}
\label{sec:egi-formulation}

The discrete EGI of a convex polytope is its surface-area measure, expressed as normal--area pairs \citep{horn1984_extended_gaussian_images}:
\begin{align}
    \mathcal{E}
    = \left\{(\mathbf{n}_i,a_i)\right\}_{i=1}^{n},
    \qquad
    \mathbf{n}_i\in\mathbb{S}^{2},
    \quad
    a_i>0,
    \label{eq:egi_set}
\end{align}
where $\mathbf{n}_i$ and $a_i$ are, respectively, the outward unit normal and area of facet $i$, and
$\mathbb{S}^{2}=\{\mathbf{x}\in\mathbb{R}^{3}:\|\mathbf{x}\|_2=1\}$. 

For a closed polytope, the facet areas and outward normals satisfy the geometric closure condition \citep{horn1984_extended_gaussian_images}
\begin{align}
    \sum_{i=1}^{n}a_i\mathbf{n}_i=\mathbf{0}.
    \label{eq:egi_closure}
\end{align}
Under this condition, and provided that the normals are not confined to a great circle, Minkowski's theorem
\citep{minkowski1989_allgemeine_lehrs_atze} states that the EGI determines a convex polytope uniquely up to translation. EGI-based light-curve inversion therefore separates area estimation from convex reconstruction
\citep{robinson2022_light_curve_inversion}.

The light-curve observation model is constructed for a prescribed set of $n_c$ candidate unit normals $\{\hat{\mathbf n}_i\}_{i=1}^{n_c}$. Under the assumed convex geometry and known, spatially uniform reflectance parameters, the brightness contribution associated with each candidate is proportional to its area, with a unit-area response determined by its normal and the observation geometry \citep{robinson2022_light_curve_inversion}. Using the Blinn--Phong BRDF {in Eq.~\eqref{eq:blinn-phong}}, these responses form the reflection matrix $\mathbf G\in\mathbb R^{m\times n_c}$ with entries
\begin{align}
    G_{ji}
    &=f_r\!\left(
        \hat{\mathbf p}_i,
        \bm{\omega}_{\ell,j},
        \bm{\omega}_{o,j}
    \right)
    \left(\bm{\omega}_{\ell,j}\cdot\hat{\mathbf n}_i\right)_+
    \left(\bm{\omega}_{o,j}\cdot\hat{\mathbf n}_i\right)_+,
    \label{eq:G_reflection_matrix}
\end{align}
where $\hat{\mathbf p}_i=\{\hat{\mathbf n}_i,C_d,C_s,s\}$ and $(x)_+=\max(x,0)$. Nonilluminated or invisible candidates contribute zero; the BRDF is evaluated only for contributing candidates. Summing the candidate contributions gives the predicted normalized light curve,
\begin{align}
    \hat{\mathbf y}=\mathbf G\mathbf a,
    \qquad
    \mathbf a\succeq\mathbf 0,
    \qquad
    \mathbf a\in\mathbb R^{n_c}.
    \label{eq:egi_linear_model}
\end{align}
With the candidate normals and reflection matrix fixed, EGI estimation consists of recovering the nonnegative area coefficients $\mathbf a$ from the measured light curve $\mathbf y$.

\subsection{Area-Image Representation}
\label{sec:area_image}

The distribution of area over normal directions depends on the object's geometry. A polytope produces a discrete EGI concentrated at its facet normals, whereas curved surfaces distribute area over a continuum of directions. To accommodate both patterns within one reconstruction framework, the Gaussian sphere is discretized into an angular grid, and the area coefficients are arranged as a two-dimensional image.

\begin{figure}[H]
    \begin{subfigure}{0.48\linewidth}
        \centering
        \includegraphics[width=0.7\linewidth]{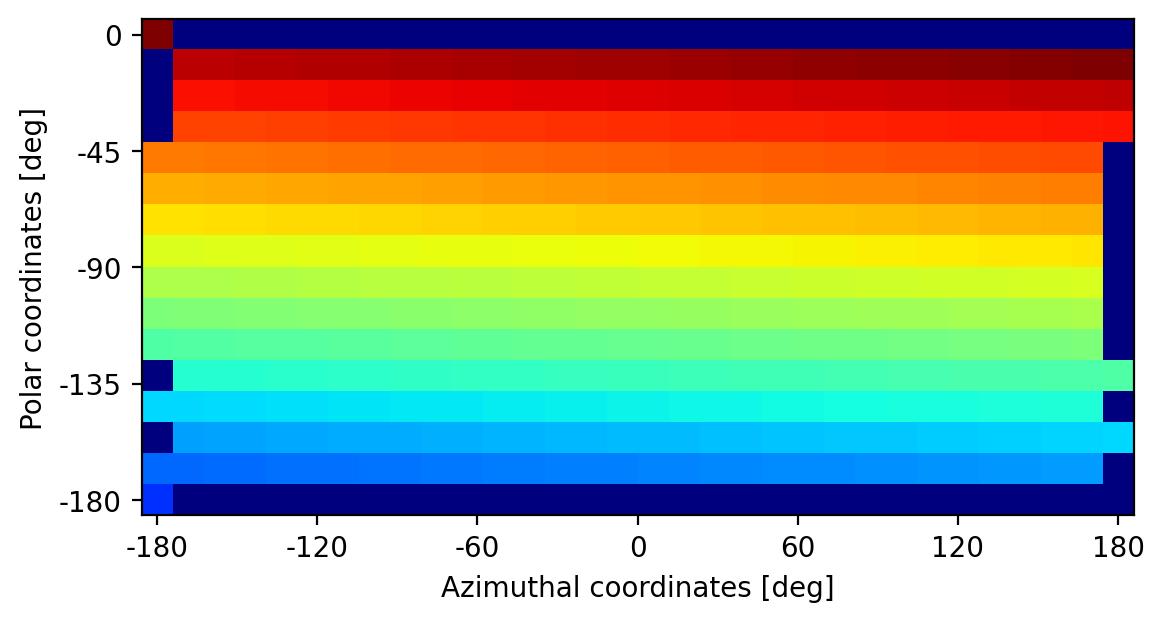}
        \caption{}
        \label{fig:masked_rectangular_grid}
    \end{subfigure}
    \begin{subfigure}{0.48\linewidth}
        \centering
        \includegraphics[width=0.4\linewidth]{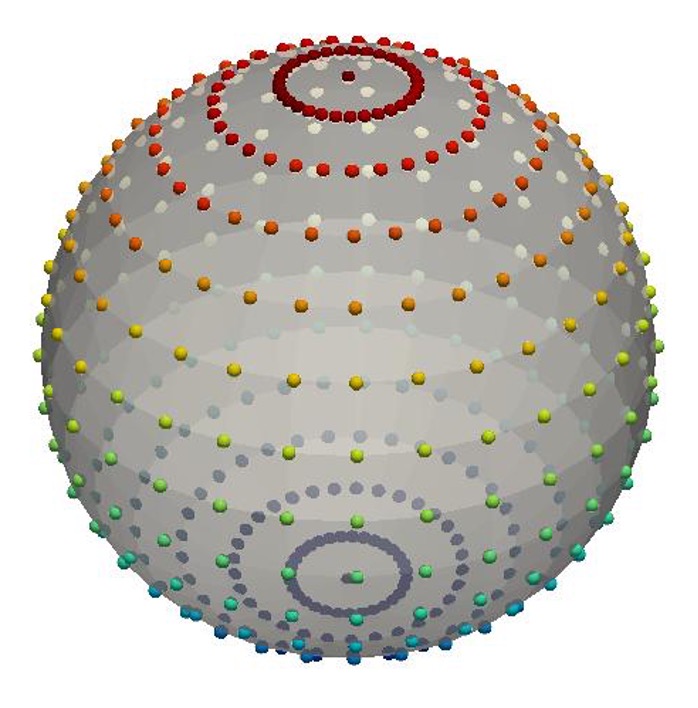}
        \caption{}
        \label{fig:normals_sphere}
    \end{subfigure}
    \caption{Example of sampling normal vectors on a $16\times32$ rectangular grid. (a) Normal samples on the masked rectangular grid, where dark blue pixels represent invalid regions (zeros in the mask). (b) Corresponding normal samples visualized as vertices on the unit sphere.}
    \label{fig:EGI_unwrapping_example}
\end{figure}

An equirectangular parametrization maps normal directions on the Gaussian sphere to image locations according to their azimuth and polar angles. Each valid pixel corresponds to a sampled normal direction. The angular grid is generated by uniformly sampling the azimuth and polar angles using $2N_\theta$ and $N_\theta$ samples over
\begin{align}
    \phi\in[-180^\circ,180^\circ], \qquad \theta\in[-180^\circ,0^\circ],
\end{align}
respectively. The grid is defined in the reconstruction coordinate frame. The $XZ$-plane is selected as the prime meridian and the $XY$-plane as the equatorial plane.

The equirectangular mapping does not preserve spherical distances or solid angles. Equal pixel separations correspond to different angular separations depending on latitude, and directions adjacent across the longitude seam appear at opposite image boundaries. The rectangular image therefore provides a coordinate-dependent representation of neighborhoods on the Gaussian sphere.

Uniform angular sampling produces many redundant sampling directions near the poles because neighboring pixels correspond to nearly identical surface normals. To improve the sampling efficiency, a spherical mesh is first constructed using the sampled grid points as vertices. Duplicate cells are then removed according to their spatial proximity, producing a binary mask whose valid pixels correspond to unique surface-normal directions. Let $M$ denote the number of valid pixels. {Taking the $M$ valid sampling directions as the candidate normals, so that $n_c=M$, the unknown EGI is represented by the coefficient vector $\mathbf a\in\mathbb R^{M}$ of Eq.~\eqref{eq:egi_linear_model},} 
whose $i$-th element stores the area assigned to the $i$-th valid normal direction. The vector contains the valid coefficients of the \emph{area image}; their pixel locations are specified by the mask. Figure~\ref{fig:EGI_unwrapping_example} illustrates the resulting binary mask together with the corresponding sampling directions on the unit sphere.

{Neighboring pixels represent nearby surface-normal directions.} This organization makes local relationships in normal space available to image-based regularization. It does not preserve the spatial adjacency of facets on the object, which is not explicitly encoded by the EGI \citep{horn1984_extended_gaussian_images}.

The reflection matrix $\mathbf{G}\in\mathbb{R}^{m\times M}$ is constructed according to Eq.~\eqref{eq:G_reflection_matrix}, with the $M$ valid sampling directions taken as the candidate normals $\hat{\mathbf{n}}_i$, so that each column represents the light-curve contribution of a unit facet area associated with one sampled normal direction. For this prescribed grid and known reflectance parameters, the observation model remains linear in the area coefficients, including both diffuse and specular contributions. Area-image estimation is therefore a linear inverse problem for which regularization supplies additional prior information. A conventional regularized formulation is
\begin{align}
    \hat{\mathbf{a}}=\arg\min_{\mathbf{a}}\frac{1}{2}\|\mathbf{y}-\mathbf{G}\mathbf{a}\|_2^2+\mathcal{P}(\mathbf{a}), \label{eq:model-based-image-recovery}
\end{align}
where $\mathbf{y}$ is the normalized light curve and $\mathcal{P}(\mathbf{a})$ represents an image regularizer.

\FloatBarrier
\subsection{{EGI Recovery with the Denoiser}} \label{sec:plug-and-play-image-recovery}
To introduce an adaptable image prior, the recovery uses PnP-ADMM \citep{venkatakrishnan2013_plug_play_priors}. The proximal step associated with an explicit regularizer is replaced by an image denoiser, while the data-fidelity step retains the physical observation operator in Eq.~\eqref{eq:egi_linear_model}.

The image reconstruction alternates between a measurement-consistency step and denoising operations. To separate these operations, an auxiliary area image $\mathbf{v}$ and a scaled dual variable $\mathbf{u}$ are introduced. At iteration $k$, the updates are
\begin{subequations}
\label{eq:ADMM-updates}
\begin{align}
    \mathbf{a}^{(k+1)}
    &= \arg\min_{\mathbf{a}} \frac{1}{2}\|\mathbf{y}-\mathbf{G}\mathbf{a}\|_2^2 + \frac{\rho_k}{2}\|\mathbf{a}-(\mathbf{v}^{(k)}-\mathbf{u}^{(k)})\|_2^2, \label{eq:ADMM-X-update}
    \\
    \mathbf{z}^{(k+1)}
    &= \Pi_{[0,1]}\left(\mathbf{a}^{(k+1)}+\mathbf{u}^{(k)} \right), \label{eq:ADMM-box-mapping}
    \\
    \mathbf{v}^{(k+1)}
    &= (1-\alpha)\mathbf{z}^{(k+1)} + \alpha\,\Pi_{[0,1]}\!\left(\mathcal{D}_{\sigma_k}(\mathbf{z}^{(k+1)})\right), \label{eq:ADMM-V-update}
    \\
    \mathbf{u}^{(k+1)} &= \mathbf{u}^{(k)} + \mathbf{a}^{(k+1)}-\mathbf{v}^{(k+1)},
    \label{eq:ADMM-U-update}
    \\
    \sigma_k &= \sqrt{\lambda/\rho_k}. \label{eq:ADMM-denoising-strength}
\end{align}
\end{subequations}
Here, $k$ is the iteration index; $\mathbf{v}^{(k)}$ is the auxiliary, denoised area-image estimate; $\mathbf{u}^{(k)}$ is the scaled dual variable; and $\rho_k>0$ is the ADMM penalty parameter. The operator $\Pi_{[0,1]}(\cdot)$ denotes element-wise clamping to the interval $[0,1]$, and $\mathbf{z}^{(k+1)}$ is the clamped input to the denoiser. The operator $\mathcal{D}_{\sigma_k}(\cdot)$ denotes an image denoiser with denoising strength $\sigma_k$, while $\lambda>0$ controls the overall denoising strength through Eq.~\eqref{eq:ADMM-denoising-strength}. The update in Eq.~\eqref{eq:ADMM-X-update} promotes measurement consistency, while Eq.~\eqref{eq:ADMM-V-update} incorporates prior information through the denoiser.

{The initial area image $\mathbf{a}^{(0)}$ is obtained by nonnegative FISTA \citep{beck2009_fast_iterative_shrinkage}, which solves an $\ell_1$-regularized nonnegative least-squares problem to obtain a sparse solution. Pixels of $\mathbf{a}^{(0)}$ exceeding $10^{-8}$ define the active support, to which the PnP-ADMM iterations are restricted. Pixels outside this support are set to zero when the area image is passed to the denoiser. The remaining variables are initialized as $\mathbf{v}^{(0)}=\mathbf{a}^{(0)}$ and $\mathbf{u}^{(0)}=\mathbf{0}$.}

The updates in Eqs.~\eqref{eq:ADMM-box-mapping} and \eqref{eq:ADMM-V-update} differ from the standard PnP update in two aspects. {First, the denoiser input and output are clamped element-wise to $[0,1]$, so that the auxiliary image, a convex combination of the two, also lies in $[0,1]$.} Second, the denoiser correction is controlled by the blending coefficient $\alpha\in[0,1]$. When $\alpha=1$, the update uses the projected denoiser output directly, whereas $0\leq\alpha<1$ damps the denoiser correction toward its input $\mathbf{z}^{(k+1)}$. In particular, $\alpha=0$ suppresses the denoiser entirely and reduces the update to box projection. The parameter $\alpha$ is distinct from the over-relaxation parameter used in some conventional ADMM formulations. Its effect is found to be significant empirically and is therefore examined in the parameter-tuning study.

The penalty parameter $\rho_k$ is adapted using a residual-balancing rule based on \citet[Section~3.4.1]{boyd2011_distributed_optimization_statistical}. After the updates in Eqs.~\eqref{eq:ADMM-X-update}--\eqref{eq:ADMM-U-update}, the primal and dual residual norms are evaluated as:
\begin{align}
    r_{\text{primal}}^{(k+1)} = \|\mathbf{a}^{(k+1)}-\mathbf{v}^{(k+1)}\|_2, \quad 
    r_{\text{dual}}^{(k+1)} = \rho_k \|\mathbf{v}^{(k+1)} - \mathbf{v}^{(k)}\|_2.
\end{align}
The penalty for the subsequent iteration is then determined by:
\begin{align}
    \rho_{k+1} =
    \begin{cases}
        \min\left(\gamma\rho_k, \ \rho_{\text{max}} \right), & \text{if} \quad r_{\text{primal}}^{(k+1)} > {\xi} \, r_{\text{dual}}^{(k+1)}, \\
        \max\left(\rho_k/\gamma, \ \rho_{\text{min}} \right), & \text{if} \quad r_{\text{dual}}^{(k+1)} > {\xi} \, r_{\text{primal}}^{(k+1)}, \\
        \rho_k, & \text{otherwise},
    \end{cases}
\end{align}
where ${\xi}>1$ is the residual-imbalance threshold 
, $\gamma>1$ is the multiplicative penalty-update factor, and $\rho_{\min}$ and $\rho_{\text{max}}$ are the prescribed lower and upper bounds on the penalty parameter, respectively. The same factor $\gamma$ is used for both increasing and decreasing $\rho_k$. When the penalty changes, the scaled dual variable is rescaled according to
\begin{align}
    \mathbf{u}^{(k+1)}
    \leftarrow
    \frac{\rho_k}{\rho_{k+1}}\mathbf{u}^{(k+1)} \label{eq:scale-dual}
\end{align}
to preserve the corresponding unscaled dual variable.

The iterations terminate when both the primal and dual residuals satisfy the absolute- and relative-tolerance criteria of {\citet{boyd2011_distributed_optimization_statistical}}{, with tolerances $\varepsilon_{\mathrm{abs}}$ and $\varepsilon_{\mathrm{rel}}$,} or when the maximum number of iterations{, ${k_{\max}}$,} is reached. {The auxiliary image $\mathbf{v}^{(k)}$ at termination is taken as the recovered area image.} A reconstruction is declared converged only when the residual criteria are satisfied. Because the denoising operator is not necessarily the proximal operator of a convex regularizer, these residual tests provide practical stopping criteria but do not establish theoretical convergence of the PnP iteration; conditions under which PnP-ADMM admits fixed-point convergence guarantees are discussed by \citet{chan2017_plug_play_admm}.

A sensitivity analysis shows that reconstruction accuracy is substantially more sensitive to the denoising parameter $\lambda$ and blending coefficient $\alpha$ than to the initial penalty $\rho_0$ and penalty-update factor $\gamma$. Accordingly, $\rho_0$ and $\gamma$ are fixed at $10^{-2}$ and $1.5$, respectively. The parameters $\lambda$ and $\alpha$ are selected using an input-dependent lookup table. The remaining parameters are held fixed throughout the experiments at ${\xi}=10.0$, $\rho_{\min}=10^{-6}$, $\rho_{\max}={1.3}$, $\varepsilon_{\mathrm{abs}}={10^{-5}}$, {$\varepsilon_{\mathrm{rel}}=10^{-5}$, and ${k_{\max}}=500$}. 

\subsection{DRUNet Denoising Prior}
\label{sec:DRUNet}
In this study, the denoising operator $\mathcal{D}_{\sigma_k}$ in Eq.~\eqref{eq:ADMM-V-update} uses the grayscale DRUNet architecture of \citet{zhang2022c_plug_play_image}. DRUNet is based on the U-Net architecture \citep{ronneberger2015_net_convolutional_networks} and consists of an encoder that progressively maps the noisy input to a lower-dimensional latent representation and a decoder that reconstructs the denoised image at the original resolution. Skip connections transmit features between encoder and decoder levels at corresponding spatial resolutions, thereby preserving fine-scale information that may otherwise be lost during downsampling. Residual blocks \citep{he2016_deep_residual_learning} are additionally used within the network to facilitate the training of the deep architecture.

Although the original DRUNet is trained using natural images, area images differ substantially from natural-image data. For example, area images are nonnegative, highly sparse, and characterized by a large dynamic range. Faceted shapes can produce highly localized peaks, whereas smooth axisymmetric surfaces can produce extended ring-like structures. Area images are also represented on an equirectangular grid, for which the azimuthal coordinate is periodic and all pixels at either pole correspond to the same physical direction. The PnP-ADMM formulation can use the pretrained denoiser without EGI-specific fine-tuning. Nevertheless, these differences motivate adapting its weights to area images. The adaptation uses a synthetic area-image dataset and a two-stage fine-tuning procedure. The reconstruction results reported in this study use the resulting fine-tuned checkpoint.


\subsection{Recovery in a Randomly Rotated Frame}
\label{sec:rotated_frame_recovery}

Area-image recovery is performed in an auxiliary reconstruction frame to vary the alignment of the object normals with the fixed angular grid. This is a coordinate transformation, not a change in the object's physical attitude. The attitude history remains assumed precisely known. Let $Q_j\in\mathrm{SO}(3)$ map world-frame coordinates into the body frame at observation time $j$, and let $R\in\mathrm{SO}(3)$ be a randomly selected, time-independent rotation from the body frame into the reconstruction frame. The illumination and viewing directions used for recovery are
\begin{align}
    \boldsymbol{\omega}_{\ell,j}^{\,\mathrm{rot}} = R\boldsymbol{\omega}_{\ell,j}=RQ_j\boldsymbol{\omega}_{\ell,j}^{\,W}, \qquad
    \boldsymbol{\omega}_{o,j}^{\,\mathrm{rot}} = R\boldsymbol{\omega}_{o,j}=RQ_j\boldsymbol{\omega}_{o,j}^{\,W},
\end{align}
where the superscript $W$ denotes world-frame directions. The original light curve is retained, and the reflection matrix is evaluated using these transformed directions and the fixed normal grid.

A common rotation preserves the inner products entering the Blinn--Phong model,
\begin{align}
    (R\mathbf n)^{\mathsf T}(R\boldsymbol{\omega})=\mathbf n^{\mathsf T}\boldsymbol{\omega}.
\end{align}
Thus, the coordinate transformation preserves the underlying photometric model. However, a fixed finite grid does not generally contain the rotated counterparts of its sampled normals, so the discretized recovery problem can depend on the selected reconstruction frame. After PnP-ADMM recovery, the estimated normals are transformed back into the body frame, while their associated areas remain unchanged:
\begin{align}
    \hat{\mathbf n}_i=R^{\mathsf T}\hat{\mathbf n}_i^{\,\mathrm{rot}}, \qquad \hat a_i=\hat a_i^{\,\mathrm{rot}}.
\end{align}

\subsection{Convex Reconstruction by Scale-Invariant Support Optimization}
\label{sec:proposed-shape-recovery}

The support-recovery stage reconstructs a convex polytope by estimating supporting-plane offsets from the recovered EGI. Given $\hat{\mathcal E}=\{(\hat{\mathbf n}_i,\hat a_i)\}_{i=1}^{n}$, the normals remain fixed while the offsets are optimized. The geometric formulation follows \citet{little1983_iterative_method_reconstructing,little1985a_extended_gaussian_images}; the numerical strategy combines a scale-invariant objective, a positive parameterization, and Adam updates followed by volume normalization. This strategy replaces the baseline's equality-constrained optimization with unconstrained parameter updates and scalar rescaling, avoiding explicit treatment of the nonlinear volume constraint and its Hessian approximation while retaining the geometric recovery objective.

The support values define the candidate polytope through the intersection of its supporting half-spaces. With $\mathbf h=[h_1,\ldots,h_n]^{\mathsf T}$ measured from an interior origin and $h_i>0$, this representation is
\begin{align}
    K(\mathbf h)=\left\{\mathbf x\in\mathbb R^3:\hat{\mathbf n}_i^{\mathsf T}\mathbf x\leq h_i,\ i=1,\ldots,n\right\}.
    \label{eq:baseline_halfspace}
\end{align}

The established polar-dual construction converts this intersection into a convex-hull problem \citep{little1983_iterative_method_reconstructing,robinson2022_light_curve_inversion}. Each supporting plane maps to a dual point, and the convex hull of these points determines the primal vertices:
\begin{align}
    \mathbf d_i&=\frac{\hat{\mathbf n}_i}{h_i}, \qquad i=1,\ldots,n, \label{eq:dual_vertex}\\
    \mathcal T_{\mathrm{dual}}(\mathbf h)&=\operatorname{ConvHull}\left(\{\mathbf d_i\}_{i=1}^{n}\right).
    \label{eq:dual_combinatorics}
\end{align}
Here, $\mathcal T_{\mathrm{dual}}$ denotes the triangles of the triangulated dual hull. For each triangle {$(i_1,i_2,i_3)$}, the corresponding primal vertex is obtained from 

\begin{align}
    {\mathbf D_{i_1i_2i_3}\mathbf x_{i_1i_2i_3}=\mathbf 1_3, \qquad \mathbf D_{i_1i_2i_3}=\begin{bmatrix}\mathbf d_{i_1}^{\mathsf T}\\ \mathbf d_{i_2}^{\mathsf T}\\ \mathbf d_{i_3}^{\mathsf T}\end{bmatrix}}.
    \label{eq:dual_to_primal}
\end{align}
The convex hull of the resulting vertices gives the candidate surface, from which its volume and facet areas are evaluated.

A scale-invariant objective separates shape estimation from the choice of reconstructed volume. The formulation builds on Little's support-based reconstruction and mixed-volume interpretation \citep{little1983_iterative_method_reconstructing,little1985a_extended_gaussian_images}. With $V(\mathbf h)=\operatorname{vol}(K(\mathbf h))$, Little's linear functional is
\begin{align}
    L(\mathbf h)=\hat{\mathbf a}^{\mathsf T}\mathbf h, \qquad \hat{\mathbf a}=[\hat a_1,\ldots,\hat a_n]^{\mathsf T}.
    \label{eq:little_linear_functional}
\end{align}
The support solver minimizes its volume-normalized form,
\begin{align}
    J(\mathbf h)=\frac{L(\mathbf h)}{V(\mathbf h)^{1/3}}.
    \label{eq:scale_invariant_support_objective}
\end{align}
Since $L({c}\mathbf h)={c}L(\mathbf h)$ and $V({c}\mathbf h)={c}^3V(\mathbf h)$ for ${c}>0$, 
 $J$ is unchanged by uniform scaling. Rescaling the supports to a prescribed target volume $V_0>0$ gives
\begin{align}
    L\left(\left[\frac{V_0}{V(\mathbf h)}\right]^{1/3}\mathbf h\right)=V_0^{1/3}J(\mathbf h).
    \label{eq:support_objective_equivalence}
\end{align}
Consequently, minimizing $J$ over positive supports selects the same shapes, up to scaling, as minimizing $L$ at fixed volume.

A positive support parameterization permits unconstrained parameter updates while keeping the polar-dual construction well defined. The supports are expressed as
\begin{align}
    \mathbf h(\boldsymbol\theta)=h_{\mathrm{floor}}\mathbf 1_n+\operatorname{softplus}(\boldsymbol\theta), \qquad h_{\mathrm{floor}}>0,
    \label{eq:support_parameterization}
\end{align}
where $\boldsymbol\theta\in\mathbb R^n$ and $\operatorname{softplus}(x)=\log(1+\exp x)$ is applied element-wise. The smooth mapping enforces $h_i>h_{\mathrm{floor}}$ for every parameter value, allowing gradient-based updates without explicit support-bound constraints. The positive floor also bounds the magnitudes of the dual points $\mathbf d_i=\hat{\mathbf n}_i/h_i$, preventing divergence as supports approach zero. The floor adds a numerical restriction to the positive-support formulation; the equivalence in Eq.~\eqref{eq:support_objective_equivalence} applies when this restriction does not exclude the normalized minimizer.

The reconstructed facet areas provide analytical gradients for support optimization without finite-difference geometry evaluations. Let $\widetilde a_i(\mathbf h)$ denote the candidate area associated with supporting plane $i$, with zero assigned to inactive planes, and let $\widetilde{\mathbf a}(\mathbf h)$ collect these areas. The standard support-volume identities give \citep{little1983_iterative_method_reconstructing,schneider2014_convex_bodies_brunn}
\begin{align}
    V(\mathbf h)&=\frac{1}{3}\widetilde{\mathbf a}(\mathbf h)^{\mathsf T}\mathbf h, \label{eq:baseline_volume_support}\\
    \nabla_{\mathbf h}L(\mathbf h)&=\hat{\mathbf a}, \qquad \nabla_{\mathbf h}V(\mathbf h)=\widetilde{\mathbf a}(\mathbf h).
    \label{eq:baseline_gradients}
\end{align}
Here, $\hat{\mathbf a}$ is the fixed recovered EGI, whereas $\widetilde{\mathbf a}(\mathbf h)$ changes with the candidate geometry. Differentiating $J$ and applying the chain rule yields
\begin{align}
    \nabla_{\mathbf h}J(\mathbf h)&=V(\mathbf h)^{-1/3}\left[\hat{\mathbf a}-\frac{L(\mathbf h)}{3V(\mathbf h)}\widetilde{\mathbf a}(\mathbf h)\right], \label{eq:scale_invariant_support_gradient}\\
    \nabla_{\boldsymbol\theta}J(\mathbf h(\boldsymbol\theta))&=\operatorname{sigmoid}(\boldsymbol\theta)\odot\nabla_{\mathbf h}J(\mathbf h),
    \label{eq:scale_invariant_parameter_gradient}
\end{align}
where $\operatorname{sigmoid}(x)=(1+\exp(-x))^{-1}$ and $\odot$ denotes element-wise multiplication. The parameters are updated using Adam \citep{kingma2015_adam_method_stochastic}.

Volume normalization fixes the reconstructed size after each parameter update without changing the scale-invariant objective. The updated polytope is reconstructed before normalization because its active planes and hull connectivity can change. With $\mathbf h^+$ denoting the updated supports and $V^+=V(\mathbf h^+)$ evaluated from that geometry, normalization follows the uniform-rescaling principle used by \citet{little1983_iterative_method_reconstructing}:
\begin{align}
    \mathbf h^{(k+1)}=\mathbf h^+\left(\frac{V_0}{V^+}\right)^{1/3}.
    \label{eq:exact_volume_projection}
\end{align}
This restores $V(\mathbf h^{(k+1)})=V_0$ up to numerical precision and leaves $J$ unchanged. Provided all normalized supports remain above $h_{\mathrm{floor}}$, the optimization parameters are recovered through
\begin{align}
    \boldsymbol\theta^{(k+1)}=\operatorname{softplus}^{-1}\left(\mathbf h^{(k+1)}-h_{\mathrm{floor}}\mathbf 1_n\right).
    \label{eq:support_inverse_parameterization}
\end{align}

Sphere-based initialization sets both the starting supports and the target volume. Following \citet{robinson2022_light_curve_inversion}, the initial supports equal the radius of a sphere whose surface area matches the recovered EGI:
\begin{align}
    \mathbf h_0=\sqrt{\frac{{A_{\mathrm{tot}}}}{4\pi}}\,\mathbf 1_n, \qquad {A_{\mathrm{tot}}}=\sum_{i=1}^{n}\hat a_i. 
    \label{eq:init_support}
\end{align}
The target volume is set to $V_0=V(\mathbf h_0)$.

The geometric evaluation is collected into the $\operatorname{PolarDual}$ operation in Algorithm~\ref{alg:polar_dual}. This operation implements the established polar-dual reconstruction and returns the candidate polytope, volume, and facet areas; it requires only the recovered normals and candidate supports.

\begin{algorithm}[t]
\caption{Polar-dual reconstruction and geometric evaluation}
\label{alg:polar_dual}
\begin{algorithmic}[1]
\Require Recovered normals $\{\hat{\mathbf n}_i\}_{i=1}^{n}$ and positive supports $\mathbf h$
\State $\mathbf d_i\gets\hat{\mathbf n}_i/h_i$ for $i=1,\ldots,n$
\State $\mathcal T_{\mathrm{dual}}\gets\operatorname{ConvHull}(\{\mathbf d_i\}_{i=1}^{n})$
\State Compute primal vertices {$\mathbf x_{i_1i_2i_3}$ for $(i_1,i_2,i_3)\in\mathcal T_{\mathrm{dual}}$} using Eq.~\eqref{eq:dual_to_primal}
\State $K\gets\operatorname{ConvHull}(\{{\mathbf x_{i_1i_2i_3}}\})$
\State Compute $\widetilde a_i$ by summing the boundary-triangle areas associated with supporting plane $i$, assigning zero to inactive planes
\State $V\gets\frac{1}{3}\widetilde{\mathbf a}^{\mathsf T}\mathbf h$
\State \Return $(K,V,\widetilde{\mathbf a})$
\end{algorithmic}
\end{algorithm}

Algorithm~\ref{alg:support_optimization} combines geometric evaluation with the parameter updates and volume normalization. Geometry is evaluated both before the gradient calculation and after the Adam update, and Adam retains its moment estimates across iterations. {In all experiments, $N_{\mathrm{iter}}=2{,}000$, the Adam learning rate is $0.1$, and $h_{\mathrm{floor}}$ is $0.15$ times the initial support value in Eq.~\eqref{eq:init_support}.} 

\begin{algorithm}[t]
\caption{Scale-invariant support optimization with analytical gradients}
\label{alg:support_optimization}
\begin{algorithmic}[1]
\Require Recovered EGI $\hat{\mathcal E}$, initial supports $\mathbf h_0$, support floor $h_{\mathrm{floor}}$, iteration count $N_{\mathrm{iter}}$, and Adam settings
\State $\mathbf h^{(0)}\gets\mathbf h_0$
\State $V_0\gets V(\mathbf h_0)$
\State $\boldsymbol\theta^{(0)}\gets\operatorname{softplus}^{-1}(\mathbf h_0-h_{\mathrm{floor}}\mathbf 1_n)$
\State Initialize Adam moment estimates
\For{$k=0,\ldots,N_{\mathrm{iter}}-1$}
    \State $(K^{(k)},V^{(k)},\widetilde{\mathbf a}^{(k)})\gets\operatorname{PolarDual}(\{\hat{\mathbf n}_i\}_{i=1}^{n},\mathbf h^{(k)})$
    \State Compute {$\nabla_{\boldsymbol\theta}J(\mathbf h^{(k)})$} using Eqs.~\eqref{eq:scale_invariant_support_gradient}--\eqref{eq:scale_invariant_parameter_gradient}
    \State $\boldsymbol\theta^{+}\gets\operatorname{AdamUpdate}(\boldsymbol\theta^{(k)},{\nabla_{\boldsymbol\theta}J(\mathbf h^{(k)})})$ 
    \State $\mathbf h^{+}\gets h_{\mathrm{floor}}\mathbf 1_n+\operatorname{softplus}(\boldsymbol\theta^{+})$
    \State $(K^{+},V^{+},\widetilde{\mathbf a}^{+})\gets\operatorname{PolarDual}(\{\hat{\mathbf n}_i\}_{i=1}^{n},\mathbf h^{+})$
    \State $\mathbf h^{(k+1)}\gets(V_0/V^{+})^{1/3}\mathbf h^{+}$
    \State $\boldsymbol\theta^{(k+1)}\gets\operatorname{softplus}^{-1}(\mathbf h^{(k+1)}-h_{\mathrm{floor}}\mathbf 1_n)$
\EndFor
\State \Return $K(\mathbf h^{(N_{\mathrm{iter}})})$
\end{algorithmic}
\end{algorithm}


\section{Baseline Method}
\label{sec:egi}

The baseline uses the observation model and support representation defined in Section~\ref{sec:proposed-method}, with nonnegative least squares (NNLS)-based EGI estimation and volume-constrained support optimization. The implementation follows \citet{robinson2023_light_curve_simulation}.

\subsection{EGI Estimation by Nonnegative Least Squares}
\label{sec:baseline-egi}

The baseline estimates nonnegative area coefficients on a Fibonacci grid of $n'$ candidate normals. The reflection matrix is constructed using Eq.~\eqref{eq:G_reflection_matrix} with $n_c=n'$, and the coefficients are obtained by
\begin{align}
    \hat{\mathbf a}=\arg\min_{\mathbf a\succeq\mathbf 0}\|\mathbf y-\mathbf G\mathbf a\|_2^2.
    \label{eq:baseline_EGI_opt}
\end{align}
The positive coefficients and their candidate normals define the initial EGI.

Local resampling and angular merging refine the candidate directions \citep{robinson2022_light_curve_inversion}. New normals are sampled within a cone of half-angle {$\psi$} about 
 each retained direction, and Eq.~\eqref{eq:baseline_EGI_opt} is solved again. Directions separated by less than $\theta_{\mathrm{merge}}$ are merged by summing their directed area vectors:
\begin{align}
    \mathbf e_{\mathcal C}=\sum_{i\in\mathcal C}a_i\mathbf n_i, \qquad a_{\mathcal C}=\|\mathbf e_{\mathcal C}\|_2, \qquad \mathbf n_{\mathcal C}=\frac{\mathbf e_{\mathcal C}}{a_{\mathcal C}}.
    \label{eq:egi_merge}
\end{align}
Closure correction then produces the refined EGI $\hat{\mathcal E}$.

\subsection{Shape Recovery by Volume-Constrained Support Optimization}
\label{sec:baseline-support}

The baseline reconstructs the refined EGI by minimizing Little's linear functional under an explicit volume constraint \citep{little1983_iterative_method_reconstructing}:
\begin{align}
    \hat{\mathbf h}=\operatorname*{arg\,min}_{\mathbf h\succ\mathbf 0}L(\mathbf h) \quad \text{subject to} \quad V(\mathbf h)=V_0.
    \label{eq:baseline_support_optimization}
\end{align}
The definitions of $L$, $V$, and the candidate polytope follow Section~\ref{sec:proposed-shape-recovery}. Equation~\eqref{eq:support_objective_equivalence} relates this constrained formulation to the scale-invariant objective.

A trust-region constrained solver enforces the volume constraint directly. Geometric evaluation uses Algorithm~\ref{alg:polar_dual}, with analytical gradients from Eq.~\eqref{eq:baseline_gradients} and the exact objective Hessian $\nabla_{\mathbf h}^{2}L=\mathbf 0$. The constraint Hessian is approximated using BFGS \citep[Chapter~6]{nocedal2006_numerical_optimization}, and a small positive support bound keeps the polar representation well defined. Initialization follows Eq.~\eqref{eq:init_support}, with $V_0=V(\mathbf h_0)$ computed from the baseline EGI.

\FloatBarrier
\section{Results}
\label{sec:results}

The proposed light-curve-based shape-recovery framework is evaluated through numerical simulations using the measurement model defined in Section~\ref{sec:lightcurve}. The evaluation covers denoiser adaptation, reconstruction accuracy across shapes and noise levels, comparison with the baseline, and the contributions of the two recovery stages.

\subsection{DRUNet Adaptation}
\label{sec:drunet_adaptation}
\subsubsection{Synthetic Area-Image Dataset}
\label{sec:synthetic_dataset}

\begin{figure}[htbp]
    \centering
    \includegraphics[width=\linewidth]{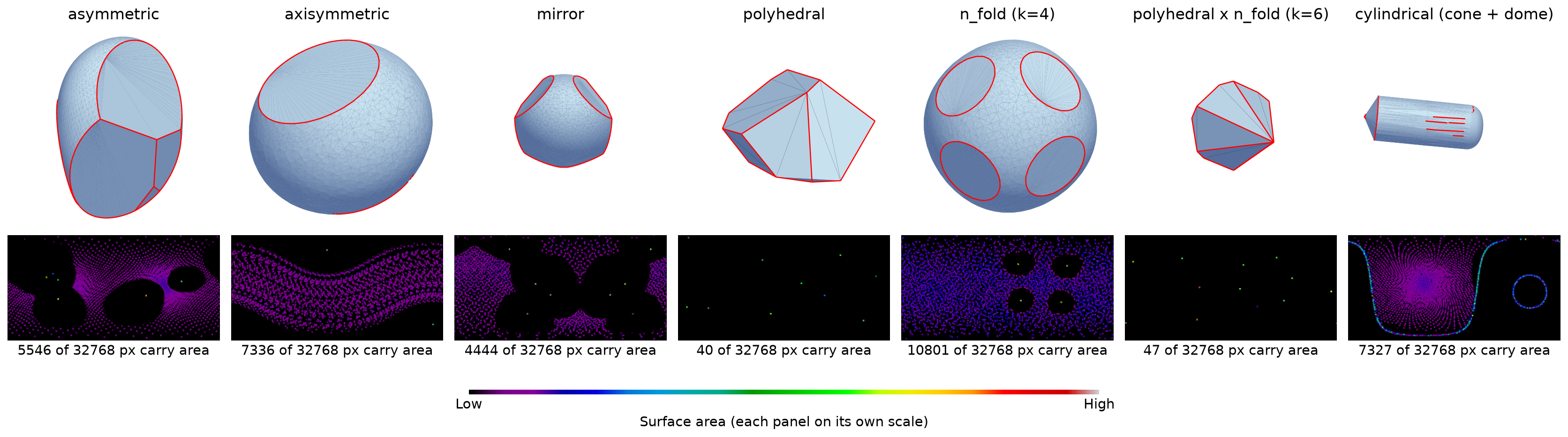}
    \caption{Representative shapes from the seven convex-shape families used to train and validate DRUNet. The top row shows the shapes, with sharp edges highlighted in red. The bottom row shows the corresponding area images after gamma correction with {exponent $0.35$} for visualization. 
}
    \label{fig:shape_families}
\end{figure}

The synthetic training dataset is generated using custom Python routines developed for this study. Each clean sample is constructed by generating a random convex mesh, computing its facet areas and outward unit normals to obtain the discrete EGI defined in Section~\ref{sec:egi-formulation}, and rasterizing the resulting surface-area distribution onto the same $128\times256$ equirectangular grid used by the reconstruction algorithm, corresponding to $N_\theta=128$ (Section~\ref{sec:area_image}). This procedure constructs area images directly from mesh geometry and does not require photometric rendering. A total of $5{,}000$ clean area images are generated, of which $4{,}486$ are assigned to the training set and $514$ to the validation set. These synthetic shapes are used only for denoiser training and parameter selection and are disjoint from the seven shape-recovery evaluation models.

The samples are drawn from seven shape families covering both asymmetric and symmetry-constrained geometries. Figure~\ref{fig:shape_families} shows a representative shape from each family with its corresponding area image. Generic asymmetric shapes represent objects without prescribed geometric symmetry. Polyhedral shapes provide an additional family characterized by a small number of distinct planar facets. The symmetry-constrained families include axisymmetric, mirror-symmetric, and three- through six-fold rotationally symmetric shapes. Symmetric polyhedral shapes combine discrete facets with prescribed symmetry. The final family consists of cylindrical shapes with flat, domed, or conical end caps. These families produce a wide range of surface-area distributions. Smooth shapes generate diffuse or ring-like structures, whereas faceted shapes produce localized peaks. Symmetric shapes introduce repeated features in the area images.

Each area image is stored as a linearly normalized 16-bit grayscale image to preserve low-amplitude features and to match the linear intensity domain encountered during reconstruction. Before quantization, the image is normalized by its maximum value. A 16-bit representation is used because area images can exhibit peak-to-median-nonzero intensity ratios of approximately $600{:}1$. An 8-bit representation would severely quantize weak features, including the ring-like structures produced by some smooth surfaces. No gamma correction or other nonlinear intensity transformation is applied because the denoiser operates on linear-valued ADMM iterates. Per-image normalization {removes the absolute area scale from the training data, so that} the denoiser learns the spatial structure of the area distribution{. During reconstruction, the ADMM iterates are passed to the denoiser without rescaling, and the denoising strength is controlled through the parameter $\lambda$ in Eq.~\eqref{eq:ADMM-denoising-strength}.}

The dataset is stratified by spatial support to prevent sparse area images from dominating individual minibatches. Spatial support is quantified using the participation ratio
\begin{equation}
S({\mathbf{a}})=\frac{\left(\sum_i {a_i}\right)^2}{\sum_i {a_i}^2}{,}
\label{eq:participation_ratio}
\end{equation}
{where $a_i$ is the value of pixel $i$ of the clean area image $\mathbf{a}$.} The participation ratio equals one when the image mass is concentrated in a single pixel and increases as the mass is distributed over more pixels. Based on this measure, the images are divided into three strata corresponding to low, intermediate, and high spatial support. Each training minibatch is constructed from a controlled mixture of these strata. Consequently, every parameter update uses area images spanning a representative range of spatial support.

\FloatBarrier
\subsubsection{Fine-Tuning Procedure and Training Behavior}
\label{sec:denoiser_training}

\begin{figure}[htbp]
    \begin{subfigure}{0.48\linewidth}
        \centering
        \includegraphics[width=\textwidth]{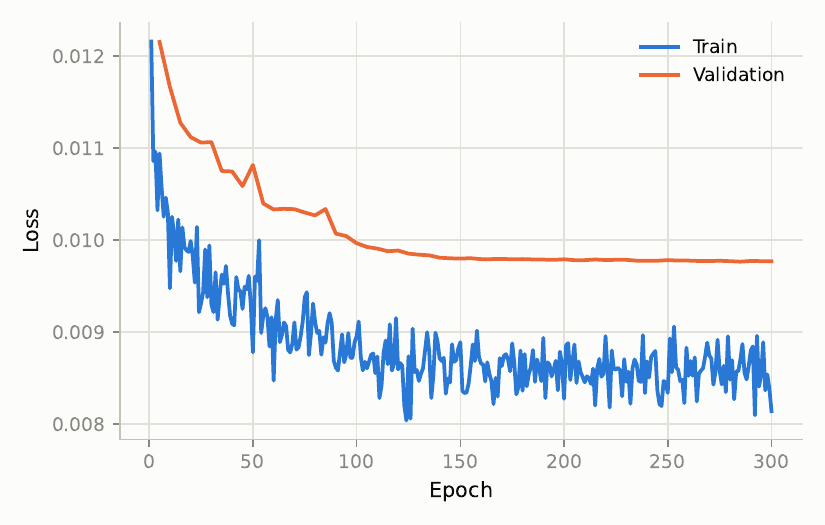}
        \caption{}
        \label{fig:loss_decay_stage1}
    \end{subfigure}
    \begin{subfigure}{0.48\linewidth}
        \centering
        \includegraphics[width=\textwidth]{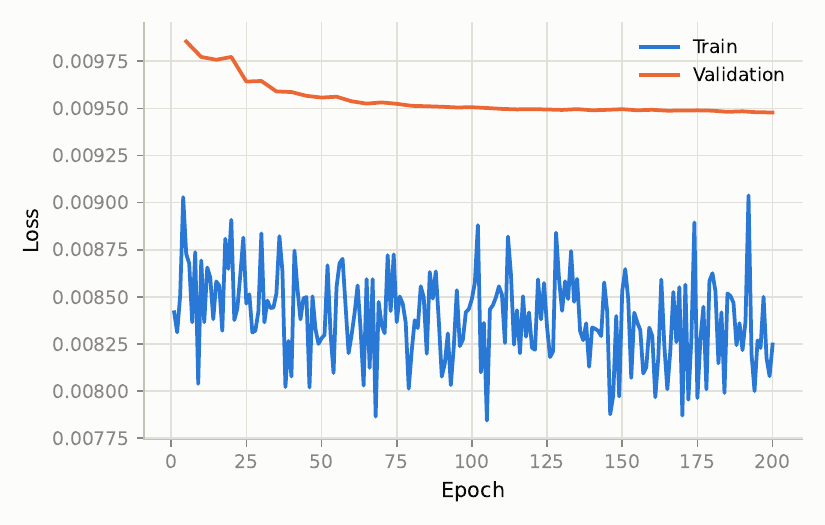}
        \caption{}
        \label{fig:loss_decay_stage2}
    \end{subfigure}
    \caption{Training and validation loss histories for the two-stage fine-tuning of DRUNet: (a) decoder-only training and (b) partial unfreezing and fine-tuning of the shallow encoder.}
    \label{fig:loss_decay}
\end{figure}

The pretrained grayscale DRUNet is adapted to area images through two-stage fine-tuning with progressively unfrozen network modules. The first stage updates only the decoder, allowing the output mapping to adapt to the sparse, high-dynamic-range structure of area images while preserving the pretrained encoder. The second stage additionally updates the input module and first downsampling module using a smaller learning rate. These shallow encoder modules capture low-level image characteristics, such as local contrast and feature scale, that differ substantially between natural and area images. The deeper encoder and bottleneck modules remain frozen throughout training.

Training pairs are synthesized on the fly from clean $96\times96$ area-image patches using a synthetic data-augmentation scheme that combines signal-dependent Poisson perturbations with additive Gaussian noise. These perturbations provide intensity-dependent variation and a background noise floor. The Gaussian-noise standard deviation $\sigma$ is supplied to the noise-conditioned DRUNet{, corresponding to $\sigma_k$ in Eq.~\eqref{eq:ADMM-denoising-strength} during reconstruction,} and is bounded above by $50/255\approx0.196$ {and by the drawn target PSNR}. A target peak signal-to-noise ratio (PSNR) is sampled uniformly between $15$ and $40~\mathrm{dB}$ to set the training corruption level. The Poisson scaling parameter is then selected using the target PSNR, the Gaussian-noise variance, and the mean intensity of the clean patch. This content-dependent calibration is intended to reduce systematic differences in corruption severity between sparse and dense patches. {During reconstruction, $\sigma_k$ remained within this training range in almost all monitored iterations.}

A foreground-weighted $\ell_1$ loss is used to prevent the large near-zero background region from dominating the training objective:
\begin{equation}
\mathcal{L}(\hat{x},x)={\beta_1}\left\|\hat{x}-x\right\|_1+{\beta_2}\left\|w\odot(\hat{x}-x)\right\|_1,\qquad w_i=1+\alpha_w\mathbb{I}(x_i>\tau),
\label{eq:training_loss}
\end{equation}
where $\hat{x}$ is the denoised output{, $x$ is the clean patch with pixel values $x_i$, $w$ is the vector of pixel weights $w_i$, $\odot$ denotes the element-wise product,} and $\mathbb{I}(\cdot)$ is the indicator function. The parameters are set to ${\beta_1=\beta_2}=1$, $\tau=0.1$, and $\alpha_w=4$. Thus, pixels with normalized intensity greater than $0.1$ receive five times the background weight in the foreground-weighted term.

During the first stage, the decoder is trained for $300$ epochs with an initial learning rate of $10^{-4}$. During the second stage, the input module and first downsampling module are additionally unfrozen and trained for $200$ epochs with {an initial} learning rate of $10^{-5}$. Both stages use the Adam optimizer and a learning-rate scheduler based on validation loss. The learning rate is reduced by a factor of $0.5$ after $10$ epochs without improvement and is bounded below by $10^{-6}$.

Figure~\ref{fig:loss_decay} shows the training and validation loss histories for both fine-tuning stages. Both losses decrease substantially during decoder-only training, with the validation loss gradually approaching a plateau. Partial encoder unfreezing produces a further reduction in the validation loss, as shown in Figure~\ref{fig:loss_decay_stage2}. The checkpoint attaining the minimum second-stage validation loss is used for all reported reconstruction results.

\FloatBarrier
\subsection{Evaluation Setup}
\label{sec:evaluation_setup}
\FloatBarrier
\subsubsection{Test Shapes and Measurement Noise}
\label{sec:test_shapes_noise}
\begin{figure}[H]
    \begin{subfigure}{0.24\linewidth}
        \centering
        \includegraphics[width=0.5\textwidth]{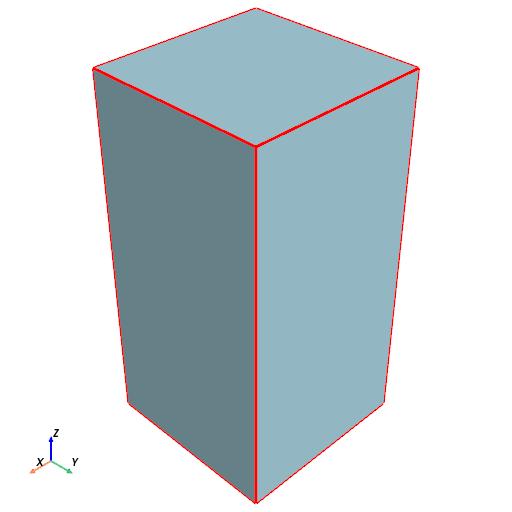}
        \caption{Box}
    \end{subfigure}
    \begin{subfigure}{0.24\linewidth}
        \centering
        \includegraphics[width=0.5\textwidth]{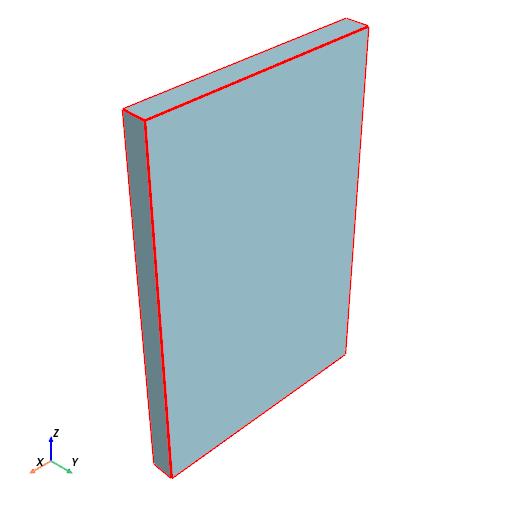}
        \caption{Flat Plate}
    \end{subfigure}
    \begin{subfigure}{0.24\linewidth}
        \centering
        \includegraphics[width=0.5\textwidth]{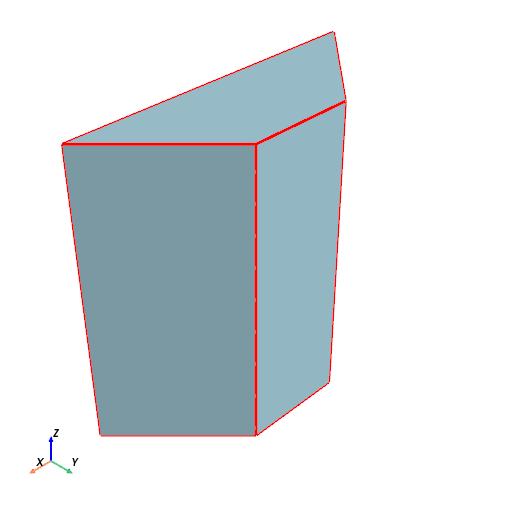}
        \caption{Trapezoidal Prism}
    \end{subfigure}
    \begin{subfigure}{0.24\linewidth}
        \centering
        \includegraphics[width=0.5\textwidth]{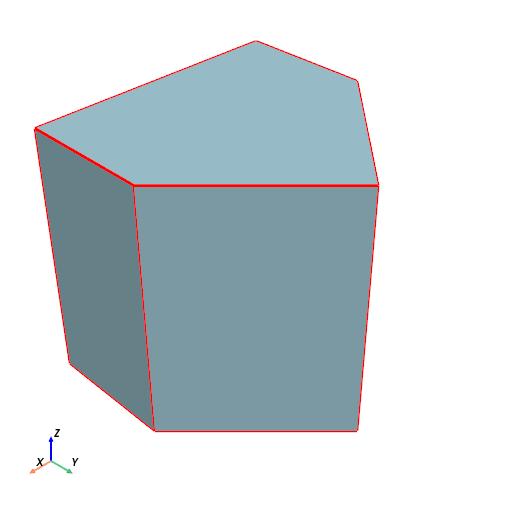}
        \caption{Pentagonal Prism}
    \end{subfigure}
    \\
    \begin{subfigure}{0.24\linewidth}
        \centering
        \includegraphics[width=0.5\textwidth]{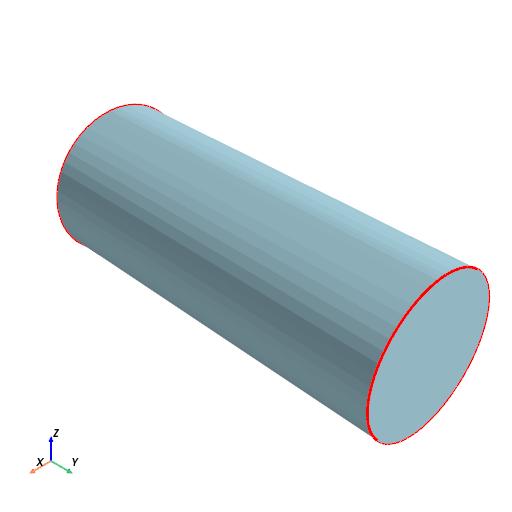}
        \caption{Cylinder}
    \end{subfigure}
    \begin{subfigure}{0.24\linewidth}
        \centering
        \includegraphics[width=0.5\textwidth]{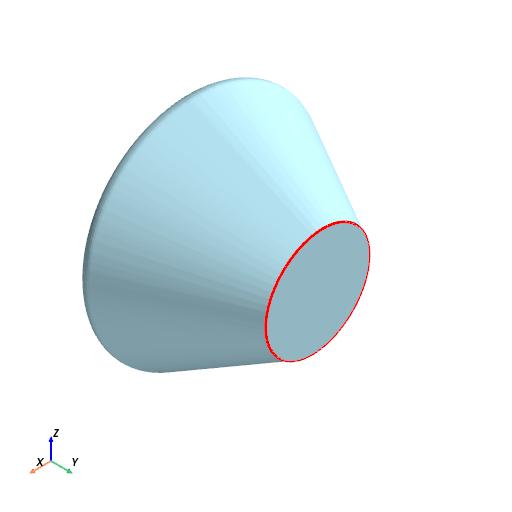}
        \caption{Capped Cone}
    \end{subfigure}
    \begin{subfigure}{0.24\linewidth}
        \centering
        \includegraphics[width=0.5\textwidth]{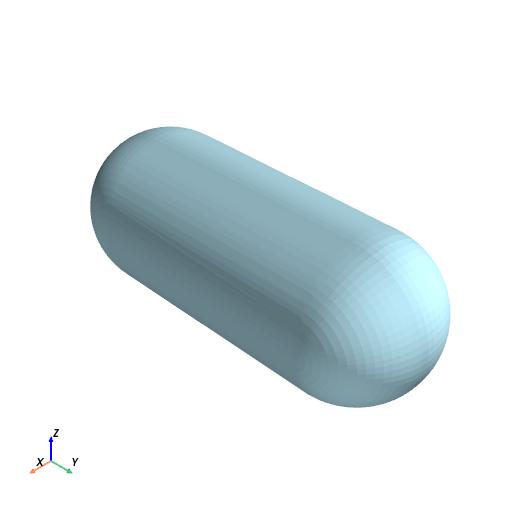}
        \caption{Capsule}
        \label{fig:capsule_true_shape}
    \end{subfigure}
    \caption{Convex models used to evaluate shape recovery. Sharp edges are highlighted in red.}
    \label{fig:convex_test_models}
\end{figure}

\begin{figure}[H]
    \begin{subfigure}{0.24\linewidth}
        \centering
        \includegraphics[width=0.85\textwidth]{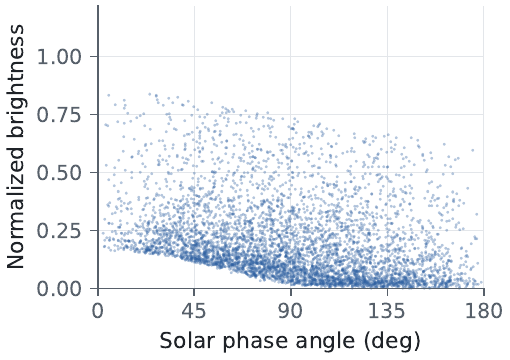}
        \caption{Box}
    \end{subfigure}
    \begin{subfigure}{0.24\linewidth}
        \centering
        \includegraphics[width=0.85\textwidth]{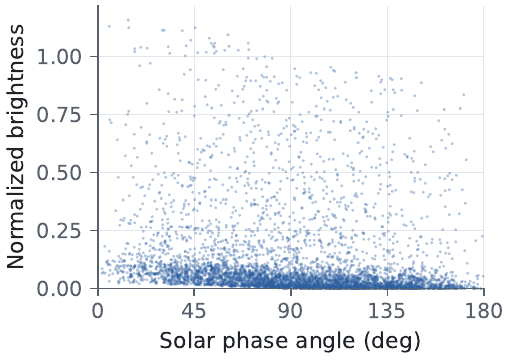}
        \caption{Flat Plate}
    \end{subfigure}
    \begin{subfigure}{0.24\linewidth}
        \centering
        \includegraphics[width=0.85\textwidth]{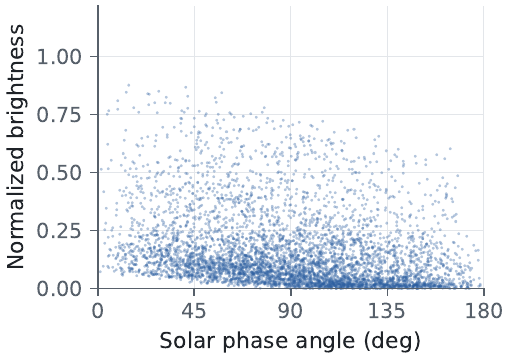}
        \caption{Trapezoidal Prism}
    \end{subfigure}
    \begin{subfigure}{0.24\linewidth}
        \centering
        \includegraphics[width=0.85\textwidth]{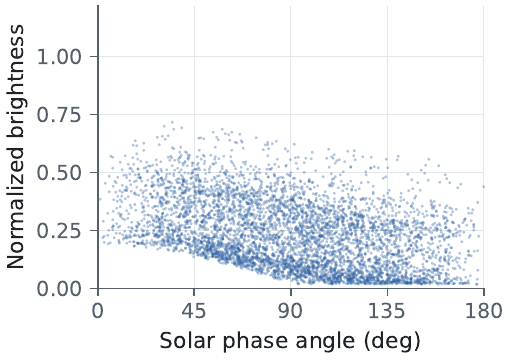}
        \caption{Pentagonal Prism}
    \end{subfigure}
    \\
    \begin{subfigure}{0.24\linewidth}
        \centering
        \includegraphics[width=0.85\textwidth]{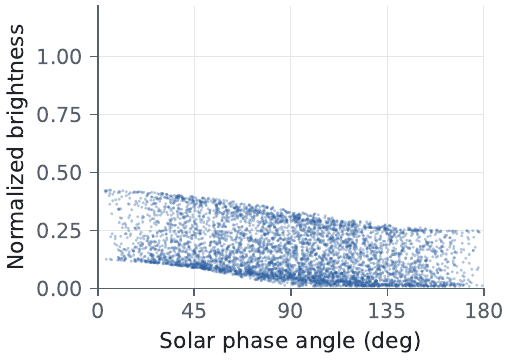}
        \caption{Cylinder}
    \end{subfigure}
    \begin{subfigure}{0.24\linewidth}
        \centering
        \includegraphics[width=0.85\textwidth]{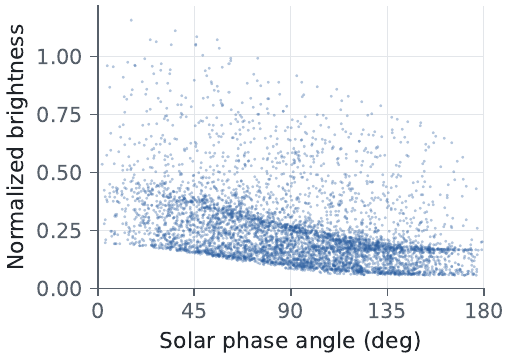}
        \caption{Capped Cone}
    \end{subfigure}
    \begin{subfigure}{0.24\linewidth}
        \centering
        \includegraphics[width=0.85\textwidth]{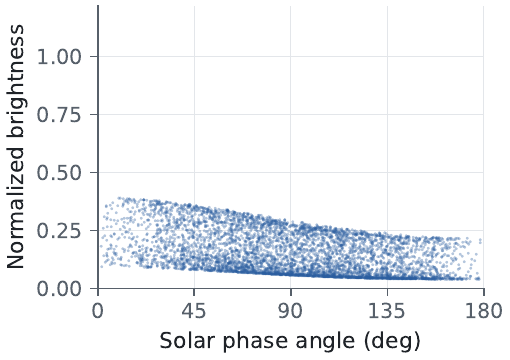}
        \caption{Capsule}
    \end{subfigure}
    \caption{Noise-free normalized brightness $y_j$ of the evaluation shapes versus the solar phase angle.}
    \label{fig:convex_light_curve}
\end{figure}

Seven convex models are used to evaluate the proposed reconstruction framework, as shown in Fig.~\ref{fig:convex_test_models}. These models are disjoint from the synthetic data used for denoiser adaptation and parameter selection (Section~\ref{sec:synthetic_dataset}). The models span two geometric regimes. The box, flat plate, trapezoidal prism, and pentagonal prism consist entirely of planar facets, producing sparse EGIs concentrated at a finite number of surface-normal directions. In contrast, the capped cone, capsule, and cylinder contain curved surfaces and therefore produce dense surface-normal distributions.

The box has edge-length ratios of $1:1:2$ and approximates a 2U CubeSat, while the flat plate represents a solar panel. The trapezoidal prism resembles spacecraft-bus geometries such as those of GRACE-FO \citep{nasa2025a_grace_mission_overview} and Sentinel-6 \citep{european2025_copernicus_sentinel_6}. The capped cone is inspired by a reentry capsule, the capsule approximates a rocket propellant tank, and the cylinder represents a rocket body. Together, these models enable evaluation across convex objects with substantially different surface-normal distributions and geometric features.

For each evaluation model, $m=5{,}000$ normalized light-curve measurements are generated using the simulation procedure described in Section~\ref{sec:lightcurve}. Illumination directions $\bm{\omega}_{\ell,j}$ and viewing directions $\bm{\omega}_{o,j}$ are sampled {independently and} uniformly from the unit sphere. {In other words, no specific attitude profile or orbit is simulated. This idealized geometry presents all orientations of the object to both the Sun and the observer. It also covers the full range of the solar phase angle $\arccos(\bm{\omega}_{\ell,j}\cdot\bm{\omega}_{o,j})$, the angle between the object-to-Sun and object-to-observer directions, from $0^\circ$ to $180^\circ$.} All models use homogeneous Blinn--Phong reflectance parameters $C_d=C_s=0.5$ and $s=10$. Four measurement-noise conditions are considered: noiseless and light-curve SNRs of $40$, $30$, and $20~\mathrm{dB}$.

{Figure~\ref{fig:convex_light_curve} shows the noise-free normalized brightness $y_j$ of the seven convex models as a function of the solar phase angle. A given phase angle corresponds to many orientations of the pair $(\bm{\omega}_{\ell,j},\bm{\omega}_{o,j})$ relative to the body frame and therefore to a range of brightness values. For the flat-faceted models, namely the box, flat plate, trapezoidal prism, and pentagonal prism, this range is wide because large flat facets produce strong specular reflection only when their normals lie near the halfway vector. For curved surfaces, such as those of the cylinder and capsule, part of the surface is always near the specular condition, so the brightness varies within a narrower band. The capped cone exhibits both behaviors.}

Unless otherwise stated, each shape and noise condition is evaluated over ten seeded reconstruction runs using the same underlying clean light curve and observation directions. Each noisy run uses a separate realization of the measurement noise described in Section~\ref{sec:lightcurve}. The proposed method additionally varies the reconstruction-frame rotation described in Section~\ref{sec:rotated_frame_recovery} using a separate random seed, while the baseline varies its normal-resampling seed. In the noiseless condition, measurement noise is omitted while the method-specific randomized operations remain active. Both methods receive identical measurements for each corresponding run. Reported means and standard deviations therefore characterize variability arising from measurement noise and method-specific randomized operations.

\subsubsection{Area-Image Evaluation Metrics}
\label{sec:area_image_metrics}
The reconstructed area image is assessed according to the agreement of its surface-area distribution with the ground truth.

Conventional raster-based metrics, such as peak signal-to-noise ratio (PSNR) and the structural similarity index (SSIM), do not directly measure angular errors in recovered area distributions. On an equirectangular grid, equal pixel weights do not account for the {$|\sin\theta|$} dependence of solid angle, where $\theta$ is the polar angle. 
 Moreover, sparse EGIs contain concentrated peaks separated by extensive near-zero regions. Raster-based comparisons can therefore be strongly influenced by background agreement and do not directly quantify the angular displacement of surface area between normal directions.

This study evaluates all recovered area images using the 1-Wasserstein distance on the unit sphere, also known as the Earth mover's distance with a geodesic transport cost \citep{kolouri2017_optimal_mass_transport}. This distance measures the minimum area-weighted angular displacement required to transform the recovered surface-normal distribution into the ground-truth distribution. Let $\{(\hat{\mathbf{n}}_i,p_i)\}_{i=1}^{N_{\mathrm{est}}}$ and {$\{(\mathbf{n}^{\mathrm{gt}}_t,q_t)\}_{t=1}^{N_{\mathrm{gt}}}$} denote the recovered and ground-truth distributions, respectively, with unit normal directions {$\hat{\mathbf{n}}_i,\mathbf{n}^{\mathrm{gt}}_t\in\mathbb{S}^2$} and normalized area weights 

\begin{equation}
{p_i=\frac{\hat a_i}{\sum_{i'=1}^{N_{\mathrm{est}}}\hat a_{i'}},
\qquad
q_t=\frac{b_t}{\sum_{t'=1}^{N_{\mathrm{gt}}}b_{t'}},}
\end{equation}
where {$\hat a_i$ and $b_t$} are the corresponding surface areas. The distance is defined as
\begin{equation}
W_1
=
{\min_{\Gamma\in\mathcal{U}(p,q)}
\sum_{i=1}^{N_{\mathrm{est}}}\sum_{t=1}^{N_{\mathrm{gt}}}
\Gamma_{it}
\arccos\!\left(\hat{\mathbf{n}}_i^{\mathsf T}\mathbf{n}^{\mathrm{gt}}_t\right)},
\label{eq:spherical_wasserstein}
\end{equation}
with the admissible transport plans
\begin{equation}
{\mathcal{U}(p,q)
=
\left\{
\Gamma\in\mathbb{R}_{\geq 0}^{N_{\mathrm{est}}\times N_{\mathrm{gt}}}:
\sum_{t=1}^{N_{\mathrm{gt}}}\Gamma_{it}=p_i,\quad
\sum_{i=1}^{N_{\mathrm{est}}}\Gamma_{it}=q_t
\right\}}.
\end{equation}
Here, $\mathbb{R}_{\geq 0}^{N_{\mathrm{est}}\times N_{\mathrm{gt}}}$ denotes the set of $N_{\mathrm{est}}\times N_{\mathrm{gt}}$ matrices with nonnegative real entries, and each entry {$\Gamma_{it}$} represents the fraction of total surface area transported from $\hat{\mathbf{n}}_i$ to {$\mathbf{n}^{\mathrm{gt}}_t$}. The transport cost is their geodesic angular separation. We report $W_1$ in degrees, with smaller values indicating better agreement.

Sinkhorn distances \citep{cuturi2013_sinkhorn} provide a computationally efficient approach to comparing distributions through entropically regularized optimal transport. In this study, this approach is adapted to EGI comparison by treating normalized surface areas as probability masses on the unit sphere and using the angular separation between surface-normal directions as the transport cost. The resulting score, termed the \emph{Sinkhorn-EGI distance}, serves as a surrogate for \(W_1\) during parameter selection. Entropic regularization favors more diffuse transport plans and enables efficient computation through iterative rescaling, known as Sinkhorn iterations, reducing the cost of repeated comparisons during parameter selection. However, regularization changes the transport plan and introduces dependence on a regularization-strength parameter, so the resulting score generally differs from the unregularized spherical 1-Wasserstein distance.

Final evaluation uses the unregularized spherical 1-Wasserstein distance to provide a common angular measure of reconstruction error without dependence on entropic regularization. The problem is solved on the full recovered EGI for both sparse and dense shapes using a network simplex algorithm \citep{bonneel2011_displacement_interpolation_lagrangian}; its computational cost is manageable for evaluation on the held-out data. Because both metrics use distributions normalized to unit mass, they measure how surface area is distributed over normal directions and are insensitive to uniform area scaling. Total-area agreement is therefore assessed separately.

\subsubsection{Shape Evaluation Metrics}
\label{sec:shape_metrics}

Reconstructed shapes are evaluated using two complementary classes of metrics: Hausdorff and Chamfer distances quantify surface discrepancies, while volumetric intersection over union (IoU) quantifies global overlap.

Following the surface-sampling procedure of \citet{cignoni1998_metro_measuring_error}, $5{,}000$ points are sampled uniformly from each reconstructed and ground-truth mesh surface. Let {$\mathcal{X}_{\mathrm{est}}$ and $\mathcal{X}_{\mathrm{gt}}$} denote the resulting point sets. The distance from a point $\mathbf{x}$ to a point set {$\mathcal{X}$} is 

\begin{equation}
{d(\mathbf{x},\mathcal{X})=\min_{\mathbf{x}'\in\mathcal{X}}\lVert\mathbf{x}-\mathbf{x}'\rVert_2}.
\end{equation}
The symmetric Hausdorff distance measures the largest sampled surface discrepancy:
\begin{equation}
{d_{\mathrm{HD}}=\max\left\{\max_{\mathbf{x}\in\mathcal{X}_{\mathrm{est}}}d(\mathbf{x},\mathcal{X}_{\mathrm{gt}}),\ \max_{\mathbf{x}\in\mathcal{X}_{\mathrm{gt}}}d(\mathbf{x},\mathcal{X}_{\mathrm{est}})\right\}}.
\label{eq:hausdorff_distance}
\end{equation}
The symmetric Chamfer distance \citep{fan2017_point_set_generation} measures the mean bidirectional surface discrepancy:
\begin{equation}
{d_{\mathrm{CD}}=\frac{1}{|\mathcal{X}_{\mathrm{est}}|}\sum_{\mathbf{x}\in\mathcal{X}_{\mathrm{est}}}d(\mathbf{x},\mathcal{X}_{\mathrm{gt}})+\frac{1}{|\mathcal{X}_{\mathrm{gt}}|}\sum_{\mathbf{x}\in\mathcal{X}_{\mathrm{gt}}}d(\mathbf{x},\mathcal{X}_{\mathrm{est}})}.
\label{eq:chamfer_distance}
\end{equation}
Unsquared Euclidean distances are used in both metrics. The distances are reported in the normalized length units of the unit-radius meshes, with smaller values indicating better surface agreement.

Global volumetric agreement is measured using the intersection-over-union (IoU) of the reconstructed and ground-truth convex bodies:
\begin{equation}
\mathrm{IoU}
=
\frac{
\operatorname{Vol}\!\left(K_{\mathrm{est}}\cap K_{\mathrm{gt}}\right)
}{
\operatorname{Vol}\!\left(K_{\mathrm{est}}\cup K_{\mathrm{gt}}\right)
},
\label{eq:volumetric_iou}
\end{equation}
where $K_{\mathrm{est}}$ and $K_{\mathrm{gt}}$ denote the convex bodies defined by the reconstructed and ground-truth meshes, respectively. The intersection and union volumes are computed directly from their convex polyhedral representations. This formulation is consistent with the convex-shape assumption adopted throughout the reconstruction framework and avoids reliance on mesh watertightness for occupancy-based volume estimation. An IoU of unity indicates perfect volumetric agreement, with lower values indicating increasing discrepancy between the reconstructed and ground-truth shapes.

\subsubsection{PnP-ADMM Parameter Selection}
\label{sec:parameter_tuning_pnp}

\begin{figure}[htbp]
    \centering
    \includegraphics[width=0.8\linewidth]{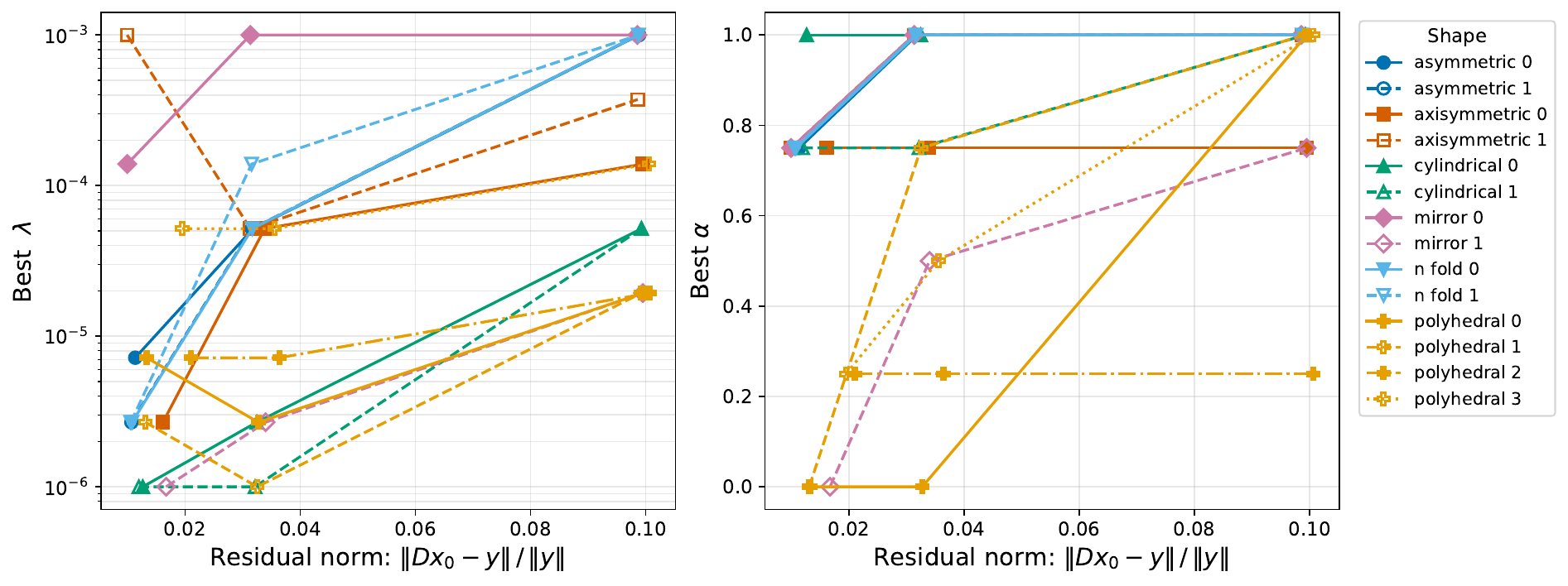}
    \caption{Grid-search-selected PnP-ADMM parameters as functions of the relative data-fit residual. Left: {denoising parameter} $\lambda$ and right: blending coefficient $\alpha$.} 
    \label{fig:lookup_table}
\end{figure}

A sensitivity analysis identifies the {denoising parameter} $\lambda$ and blending coefficient $\alpha$ as the parameters having the greatest influence on reconstruction accuracy. A grid search over $\lambda\in[10^{-6},10^{-3}]$ and $\alpha\in[0,1]$ further shows that the preferred values vary with both shape and noise level. Therefore, these parameters are selected using an input-dependent lookup table rather than a single fixed configuration.

Figure~\ref{fig:lookup_table} illustrates how the selected parameters vary with the relative data-fit residual. Larger residuals generally favor {larger blending coefficients, that is, a stronger denoiser contribution}, 
 whereas the preferred {denoising parameter} exhibits greater shape-dependent variation. Substantial differences can occur even among shapes from the same family. This behavior is particularly evident in the blending coefficients selected for the polyhedral shapes. Their area images contain a small number of highly concentrated peaks and are therefore especially sensitive to oversmoothing by the denoiser. To capture both noise level and shape-dependent structure, the lookup table uses three features computed from the initial {FISTA estimate $\mathbf a^{(0)}$}: the relative data-fit residual {$\|\mathbf G\mathbf a^{(0)}-\mathbf y\|_2/\|\mathbf y\|_2$}, {the} support fraction{, defined as the fraction of image pixels in the active support,} and {the} spatial roughness{, defined as the mean absolute gradient of the sum-normalized initial area image over its support}. 
 For each reconstruction, $\lambda$ and $\alpha$ are obtained from the nearest entry in the standardized three-dimensional feature space.

The lookup table is constructed from 14 synthetic shapes drawn from the families of Section~\ref{sec:synthetic_dataset}, evaluated at three noise conditions each, and is validated by leave-one-shape-out cross-validation {\citep{roberts2017cross}}. 
 In each fold, one shape and all three of its noise conditions are excluded from the lookup table. Parameters for the excluded shape are selected using only entries from the remaining 13 shapes, and their performance is evaluated using the corresponding Sinkhorn-EGI scores recorded during the grid search. The same entropically regularized transport objective, defined in Section~\ref{sec:area_image_metrics}, is therefore used for both grid-search optimization and held-out evaluation. Repeating this procedure for all 14 shapes evaluates the lookup strategy on shapes excluded from its construction. The resulting mean Sinkhorn-EGI score is \(1.859^\circ\), compared with \(2.362^\circ\) for a single fixed configuration selected globally to minimize the worst-case score across all 42 conditions. Input-dependent selection reduces the mean regularized score by \(21.3\%\) and outperforms the fixed configuration in 34 of the 42 held-out conditions. None of these 14 shapes is among the seven evaluation models of Section~\ref{sec:test_shapes_noise}.

\FloatBarrier
\subsection{Recovery Across Shapes and Noise Levels}
\label{sec:recovery_shapes_noise}
\subsubsection{Area-Image Recovery}
\label{sec:result_image_recovery}

Area images are recovered using the proposed method for all test shapes and light-curve noise levels. Recovery accuracy is evaluated using the 1-Wasserstein distance ($W_1$) defined in Section~\ref{sec:area_image_metrics}.

\begin{figure}[htbp]
    \centering
    \includegraphics[width=\linewidth]{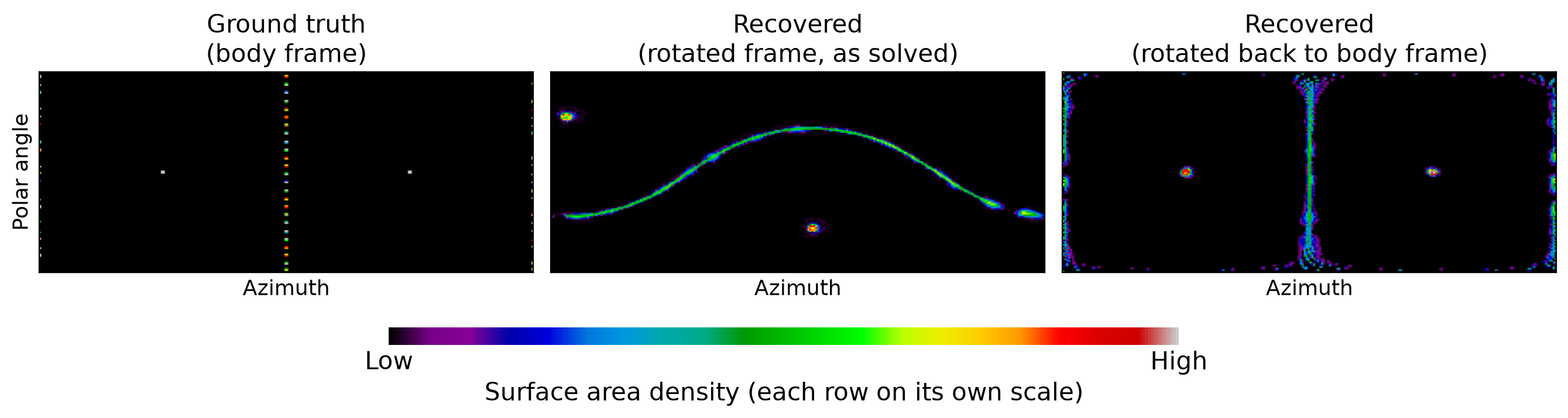}
    \caption{Area-image recovery for a cylinder model in a randomly rotated coordinate frame.}
    \label{fig:rotation_frames}
\end{figure}

Figure~\ref{fig:rotation_frames} illustrates area-image recovery using the randomly rotated frame described in Section~\ref{sec:rotated_frame_recovery}. The ground-truth area image is shown in the body frame, whereas the reconstruction is initially obtained in the randomly rotated frame. For the cylinder, the two concentrated peaks correspond to the flat end caps, while the continuous band represents the distribution of surface normals over the lateral surface. Transforming the recovered EGI back to the body frame and rasterizing it onto the original angular grid produces the image shown on the right, enabling direct visual comparison with the ground truth.

Random-frame recovery offers a practical way to mitigate unfavorable alignment between the EGI and the equirectangular grid. In a preliminary noiseless flat-plate test, four random rotations reduced the mean $W_1$ from $2.76^\circ$ for direct body-frame recovery to $2.32^\circ$ and increased the mean IoU from $0.659$ to $0.714$. Each rotation improved both metrics relative to body-frame recovery. Although limited to one geometry and noise condition, these results demonstrate that changing the reconstruction frame can improve recovery accuracy and provide empirical motivation for random rotation.

\begin{figure*}[htbp]
    \centering
    \includegraphics[width=\linewidth]{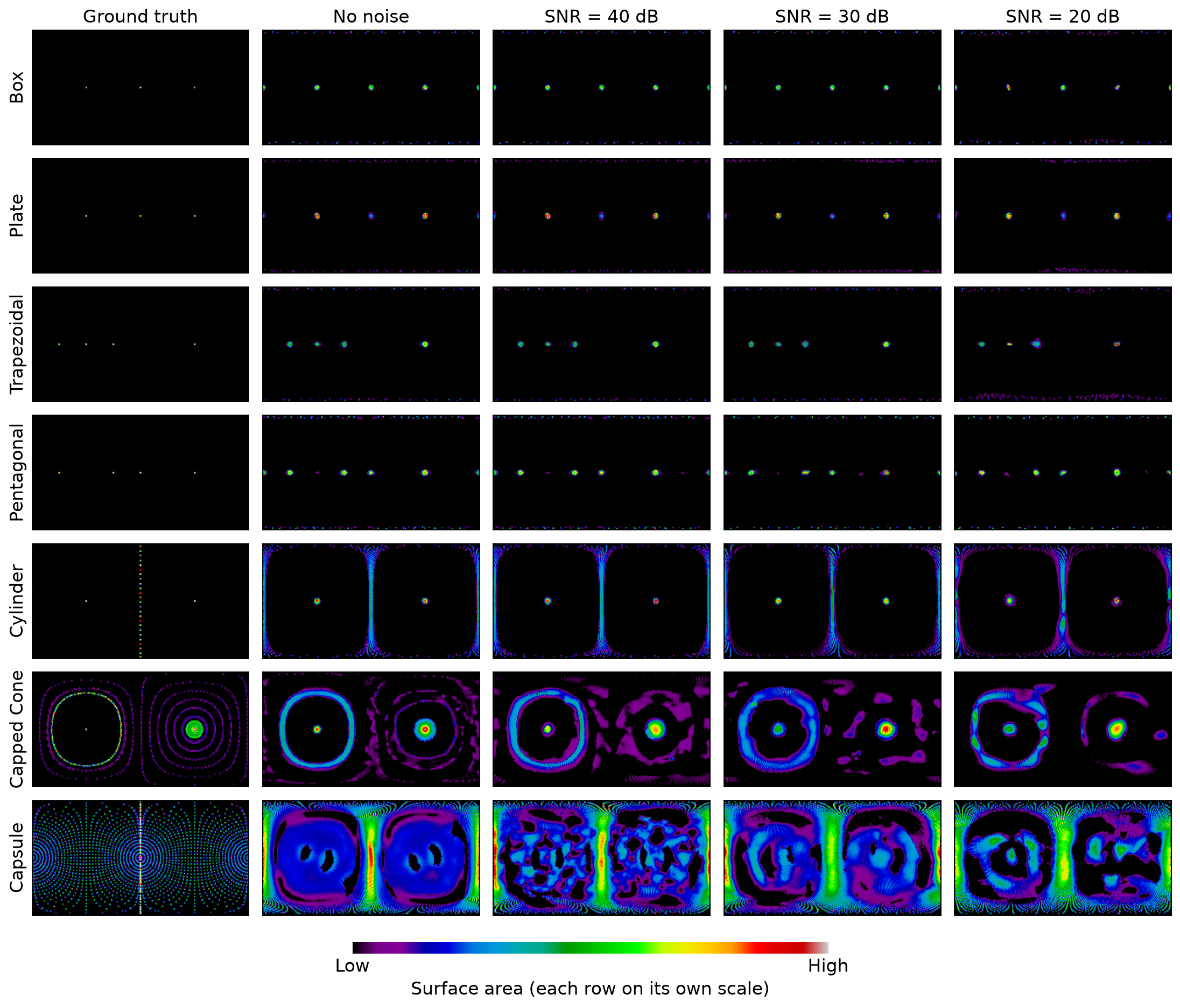}
    \caption{Recovered area images by shape, with ground truth followed by decreasing SNR. Each recovery shows the median-$W_1$ run among ten independent runs, mapped back and re-rasterized in the body frame.}
    \label{fig:area_image_grid}
\end{figure*}

\begin{figure}[htbp]
    \centering
    \includegraphics[width=0.8\linewidth]{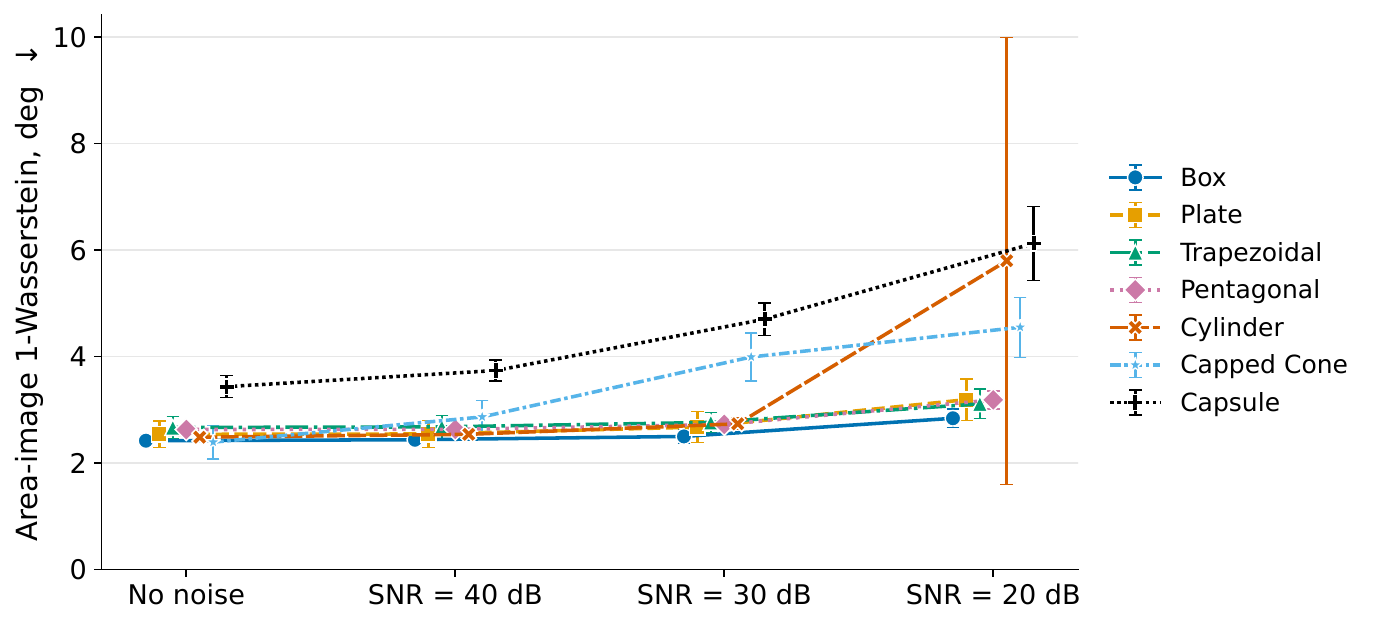}
    \caption{Area-image-recovery accuracy across shapes and noise levels over 10 reconstruction runs. Each marker represents the mean across runs, and error bars indicate $\pm1$ standard deviation.}
    \label{fig:metric_trend_wasserstein}
\end{figure}

Figure~\ref{fig:area_image_grid} compares the recovered area images across the tested shapes and noise levels, while Figure~\ref{fig:metric_trend_wasserstein} summarizes the corresponding $W_1$ accuracy over the 10 reconstruction runs. Area-image recovery remains relatively stable from the noiseless condition through moderate measurement noise, particularly for the faceted shapes. The box, flat plate, trapezoidal prism, and pentagonal prism exhibit only gradual increases in $W_1$ as the SNR decreases. For example, the mean $W_1$ for the box increases from $2.417^\circ$ in the noiseless case to $2.497^\circ$ at $30~\mathrm{dB}$ and $2.838^\circ$ at $20~\mathrm{dB}$.

The recovered area images generally exhibit broader peaks than their ground-truth counterparts. This smoothing is present even in the noiseless case and persists across the tested PnP-ADMM parameter settings, indicating that it is not caused solely by measurement noise. Instead, it is consistent with the combined effects of the ill-conditioned inverse problem, angular discretization, and regularization imposed during area-image recovery. Increasing measurement noise generally accentuates the spreading and distortion of the recovered surface-area distribution.

As shown in Fig.~\ref{fig:metric_trend_wasserstein}, the effect of measurement noise is more pronounced for shapes containing curved surfaces. The capsule exhibits the clearest systematic degradation, with its mean $W_1$ increasing from $3.432^\circ$ in the noiseless case to $4.698^\circ$ at $30~\mathrm{dB}$ and $6.122^\circ$ at $20~\mathrm{dB}$. The capped cone exhibits a similar, although less severe, increase. In contrast to the relatively flat trends of the faceted shapes, these results indicate greater sensitivity to measurement noise when the EGI contains a dense distribution of surface normals.

The cylinder exhibits a distinct behavior under the most severe noise condition. At $20~\mathrm{dB}$, its mean $W_1$ increases to $5.796^\circ$, with a standard deviation of $4.196^\circ$. The unusually large error bar in Fig.~\ref{fig:metric_trend_wasserstein} reflects intermittent failures in a small number of random-frame runs rather than uniform degradation across all runs. Additional investigation showed that these failures are sensitive to the orientation of the cylinder relative to the equirectangular grid and are associated with cases in which its ring of lateral surface normals approaches the polar regions of the parameterization. However, pole crossing alone does not fully explain the failures. The large run-to-run variation therefore indicates an orientation-sensitive failure mode of the current area-image recovery under severe measurement noise rather than a deterministic failure for cylindrical geometry.

A systematic alignment strategy was investigated to reduce this orientation sensitivity, since random rotation does not guarantee a favorable reconstruction frame. Using the initial FISTA estimate, an area-weighted normal second-moment matrix was computed, and the eigenvector associated with its smallest eigenvalue was aligned with the polar axis. This alignment minimizes the area-weighted squared projection of the estimated normals onto that axis and was intended to reduce the concentration of dominant area-image features near the poles. Although it substantially reduced run-to-run variability for the cylinder at $20\,\mathrm{dB}$, it did not materially improve the mean shape IoU and reduced it for several other shapes. These results indicate that this alignment criterion does not consistently improve reconstruction accuracy, despite reducing variability in the cylinder case. Random-frame recovery is therefore retained as a practical choice, while systematic optimization of the reconstruction frame remains beyond the scope of this study.

\FloatBarrier
\subsubsection{Shape Recovery}
\label{sec:result_shape_recovery}

\begin{figure*}[htbp]
    \centering
    \includegraphics[width=\linewidth]{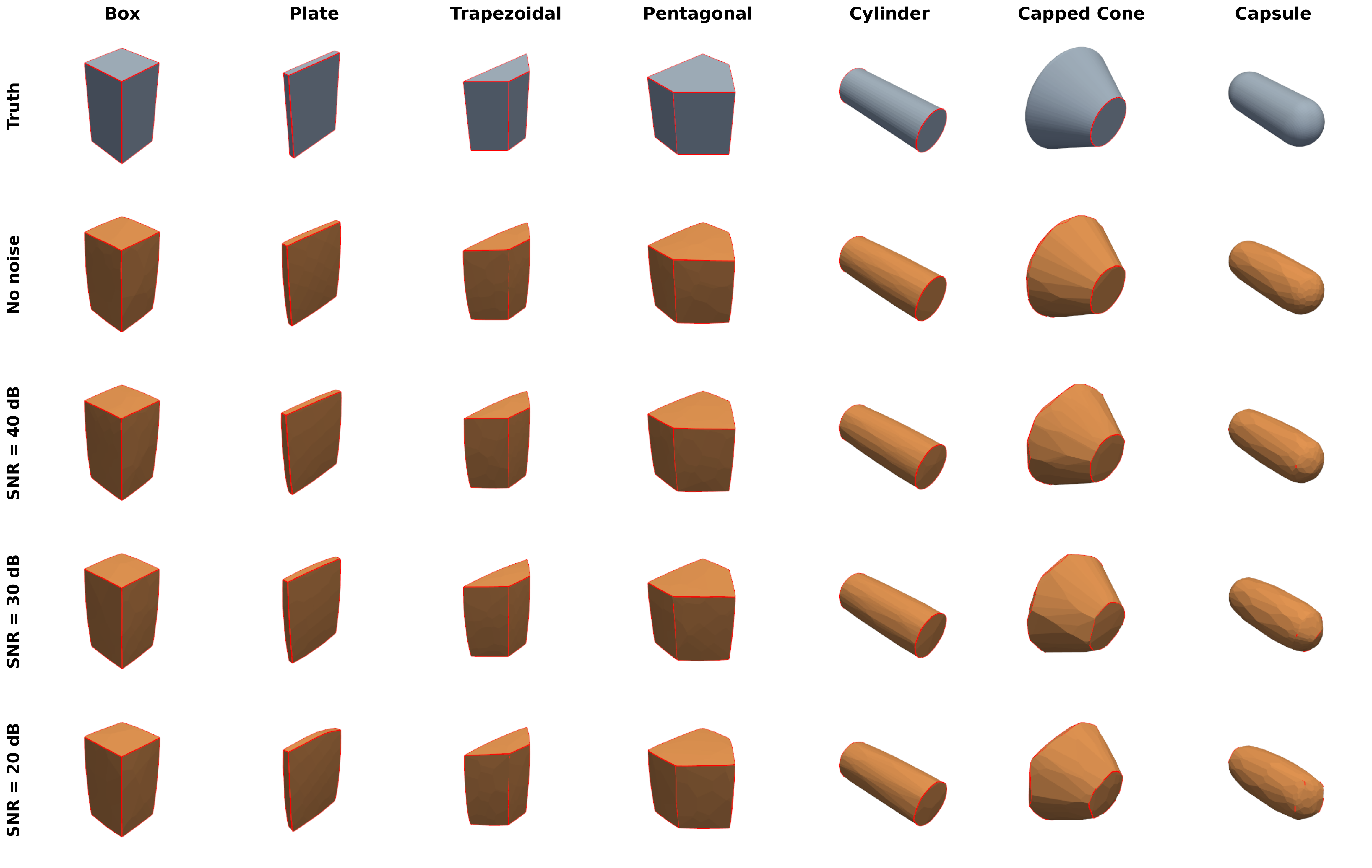}
    \caption{Ground-truth and recovered shapes across noise levels. For each shape and method, the panel shows the reconstruction corresponding
to the median IoU among the 10 reconstruction runs.}
    \label{fig:noise_sweep_grid}
\end{figure*}

\begin{figure*}[htbp]
    \centering

    \begin{subfigure}{0.8\linewidth}
        \centering
        \includegraphics[width=\linewidth]{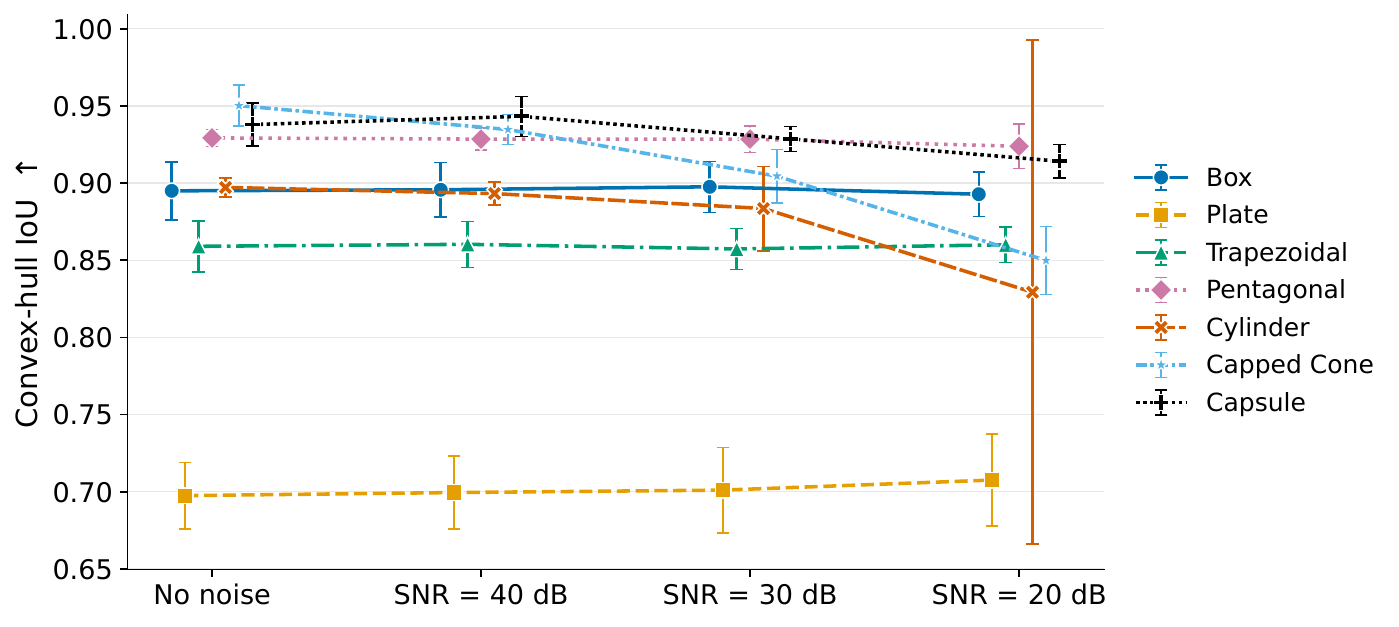}
        \subcaption{IoU}
        \label{fig:metric_trend_iou}
    \end{subfigure}

    \medskip

    \begin{subfigure}{0.8\linewidth}
        \centering
        \includegraphics[width=\linewidth]{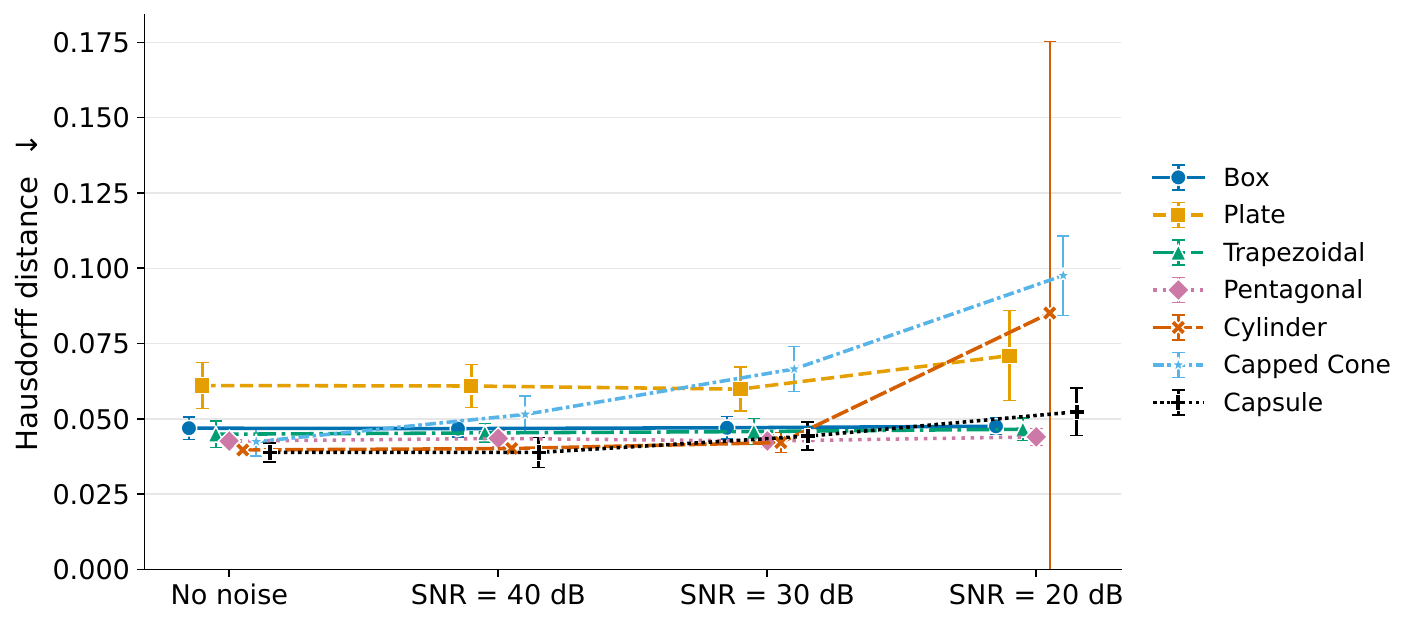}
        \subcaption{Hausdorff distance}
        \label{fig:metric_trend_hausdorff}
    \end{subfigure}

    \medskip

    \begin{subfigure}{0.8\linewidth}
        \centering
        \includegraphics[width=\linewidth]{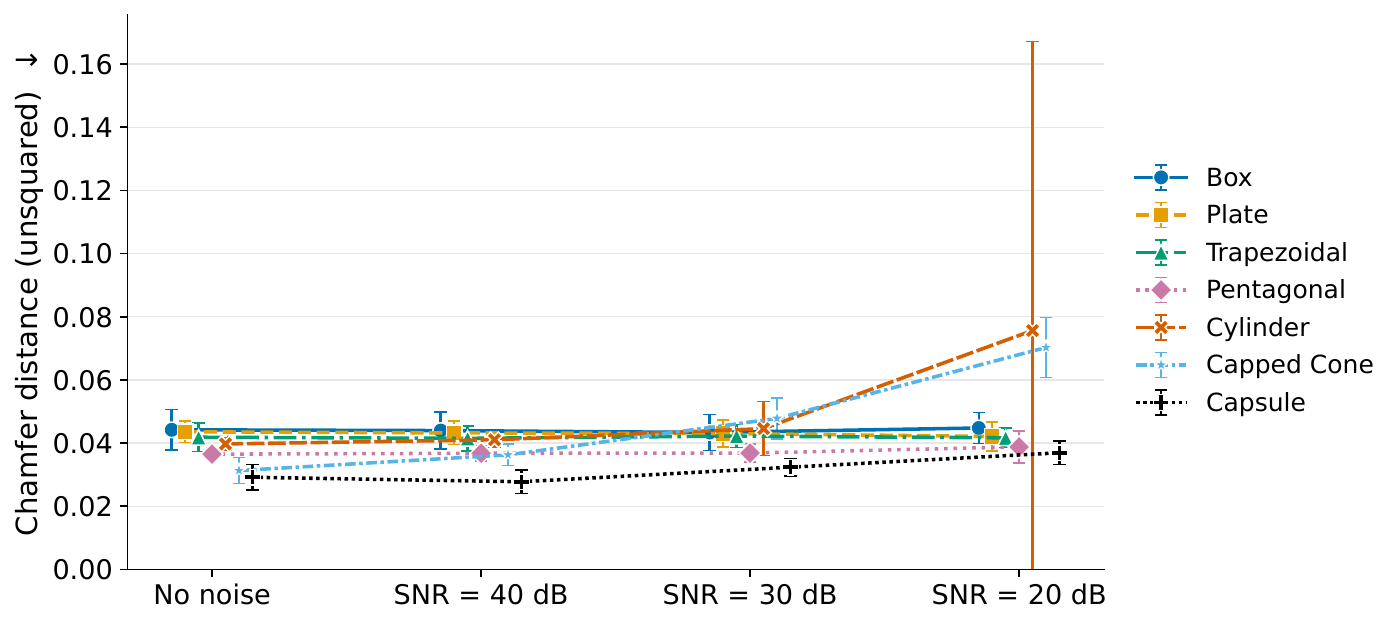}
        \subcaption{Chamfer distance}
        \label{fig:metric_trend_chamfer}
    \end{subfigure}

    \caption{Shape-recovery accuracy across shapes and noise levels over 10 reconstruction runs. Each marker represents the mean across runs, and error bars indicate $\pm 1$ standard deviation.}
    \label{fig:metric_trend_all}
\end{figure*}

Figure~\ref{fig:noise_sweep_grid} shows representative recovered shapes across the tested noise levels, while Fig.~\ref{fig:metric_trend_all} summarizes the corresponding IoU, Hausdorff distance, and Chamfer distance over the 10 reconstruction runs. Overall, the reconstructed shapes remain stable through moderate measurement noise. The mean IoU across the seven shapes decreases only from $0.881$ in the noiseless condition to $0.879$ at $40~\mathrm{dB}$ and $0.872$ at $30~\mathrm{dB}$, before decreasing more noticeably to $0.854$ at $20~\mathrm{dB}$. The mean Hausdorff distance correspondingly increases from $0.045$ to $0.064$, while the mean Chamfer distance increases from $0.038$ to $0.050$ over the same range. Thus, all three metrics indicate that most of the degradation occurs under the most severe noise condition.

The effect of measurement noise again depends on target geometry. The box, flat plate, trapezoidal prism, and pentagonal prism exhibit little systematic degradation across the tested noise levels. In contrast, the curved geometries show greater sensitivity to noise, particularly the capped cone and cylinder. For the capped cone, the mean IoU decreases from $0.950$ in the noiseless case to $0.850$ at $20~\mathrm{dB}$, accompanied by increases in both Hausdorff and Chamfer distances. The capsule is comparatively robust in the final shape domain, maintaining a mean IoU above $0.91$ even at $20~\mathrm{dB}$ despite the substantially larger $W_1$ observed for its recovered area image.

\begin{figure*}[htbp]
    \centering
    \includegraphics[width=0.8\linewidth]{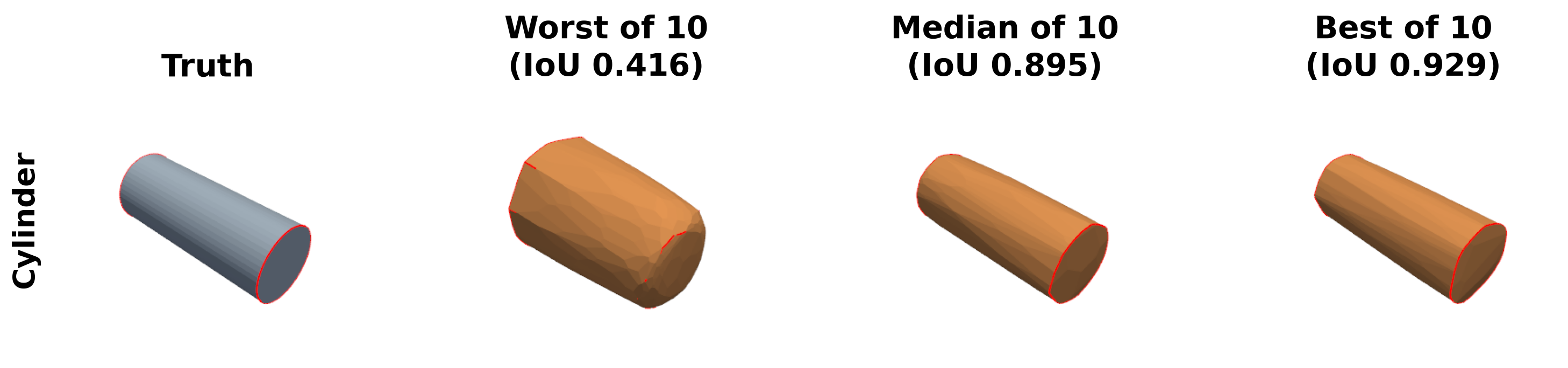}
    \caption{Run-to-run variation in the recovered cylinder shape at $20~\mathrm{dB}$.}
    \label{fig:spread_cylinder_20}
\end{figure*}

This difference between area-image and shape-recovery metrics indicates that errors in the recovered area image do not translate directly into equivalent errors in the reconstructed shape. In particular, the capsule and capped cone retain relatively high shape-recovery accuracy despite larger $W_1$ values than the faceted shapes. The support-optimization stage therefore exhibits some tolerance to distributed errors in the recovered EGI. Conversely, the cylinder at $20~\mathrm{dB}$ demonstrates that sufficiently large area-image errors can propagate to the final reconstruction. The same runs responsible for the large $W_1$ variability discussed above also produce substantially larger variation in the shape metrics, yielding an IoU of $0.829 \pm 0.163$. Figure~\ref{fig:spread_cylinder_20} illustrates this run-to-run variation. The degraded reconstructions exhibit a wider and shorter barrel-like geometry, consistent with distortion of the recovered distribution of lateral surface normals. The same axial shortening was reported by \citet{vallverducabrera2021_inversion_shape_space} for cylinder recovery from a sparse EGI, where it appeared in most runs and was attributed to the information lost in representing a lateral surface that is the limit of infinitely many infinitesimal faces. Here it appears only intermittently and only at the lowest SNR, the cylinder otherwise remaining at a mean IoU of 0.883 or above. Thus, the cylinder variability originates upstream in area-image recovery rather than representing an independent failure of the support-optimization stage.

The flat plate provides an important example of the complementary information conveyed by the three shape metrics. It has the lowest IoU among the tested shapes, remaining near $0.70$ across all noise levels, despite exhibiting relatively small $W_1$ and surface-distance errors. Its Hausdorff distance is approximately $0.06$--$0.07$, while its Chamfer distance remains approximately $0.04$, comparable to those of the other successfully reconstructed shapes. The low IoU therefore does not indicate a proportionally larger surface-reconstruction error. Rather, because the flat plate has a substantially larger aspect ratio than the other test shapes, small normal displacements of its large faces produce a relatively large change in enclosed volume and consequently a stronger reduction in volumetric IoU. The surface-distance metrics are less sensitive to this thin-body effect and indicate that the geometric reconstruction error of the flat plate is comparable to that of the other faceted shapes. This result highlights the importance of considering both volumetric-overlap and surface-distance metrics when comparing reconstruction accuracy across geometries with substantially different aspect ratios.

In summary, the shape-recovery results demonstrate that the proposed framework maintains accurate reconstruction across a broad range of target geometries and remains robust through moderate measurement noise. The degradation observed at the lowest SNR is concentrated primarily in targets with densely distributed surface normals, while the faceted shapes remain comparatively insensitive to noise. The differences among IoU, Hausdorff distance, and Chamfer distance further show that no single metric fully characterizes reconstruction quality across geometrically dissimilar targets.

\FloatBarrier
\subsection{Comparison with the Baseline}
\label{sec:result_proposed_vs_baseline}

The proposed framework is compared with the baseline using identical light-curve measurements and a common shape-evaluation procedure. Both methods use the known illumination and observation geometries, and their reconstructions are evaluated using the metrics defined in Section~\ref{sec:shape_metrics}. The baseline incorporates the corrected volume evaluation and analytical gradients described in Section~\ref{sec:baseline-support}. Each shape and noise condition is evaluated over 10 reconstruction runs.

\begin{figure}[htbp]
    \centering
    \includegraphics[width=\linewidth]{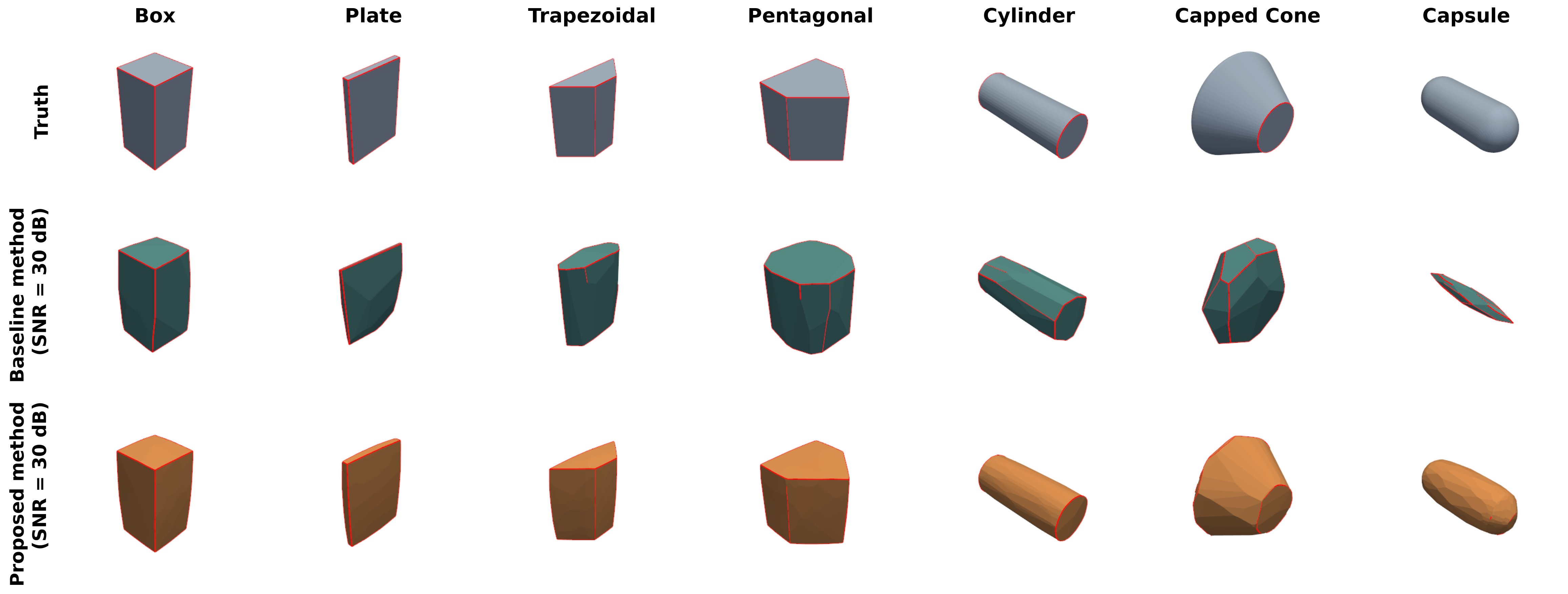}
    \caption{Qualitative comparison with the baseline method at $30~\mathrm{dB}$. For each shape and method, the panel shows the reconstruction corresponding to the median IoU among the 10 reconstruction runs.}
    \label{fig:baseline_compare_grid}
\end{figure}

The qualitative comparison at $30\,\mathrm{dB}$ illustrates the geometric differences between the two methods. Figure~\ref{fig:baseline_compare_grid} shows a representative reconstruction selected by median IoU from the 10 runs for each shape and method. Across the selected examples, the proposed method more closely reproduces the ground-truth geometry, while the baseline exhibits larger deviations in shape and proportions. These representative reconstructions complement the run-averaged metrics but do not convey the full variability within each case.

The quantitative results show substantial average improvements, although the proposed method does not outperform the baseline in every comparison. At $30\,\mathrm{dB}$, Table~\ref{tab:baseline_comparison_snr30} shows that the mean IoU across the seven shapes increases from $0.615$ to $0.872$, while the mean Hausdorff and Chamfer distances decrease from $0.198$ to $0.050$ and from $0.167$ to $0.041$, respectively. The proposed method achieves better mean values for all three metrics on six of the seven shapes. The exception is the box, for which the baseline has slightly higher mean IoU ($0.899$ versus $0.898$) and slightly lower mean Chamfer distance ($0.042$ versus $0.043$), while the proposed method has lower mean Hausdorff distance ($0.047$ versus $0.078$). Thus, performance on the box is comparable by volume overlap and average surface distance, with a smaller sampled worst-case surface error for the proposed method.

The aggregate IoU advantage persists across the tested noise conditions. As shown in Table~\ref{tab:baseline_comparison_noise}, the proposed method achieves mean IoUs of $0.881$, $0.879$, and $0.854$ under noiseless, $40\,\mathrm{dB}$, and $20\,\mathrm{dB}$ conditions, respectively, compared with $0.614$, $0.616$, and $0.613$ for the baseline. At each condition, the proposed method has higher mean IoU on all shapes except the box. The baseline's nearly constant aggregate IoU indicates that its reconstruction error over the tested range is less sensitive to measurement noise than to other limitations of the recovery pipeline.

The faceted targets demonstrate that a small number of true surface normals does not by itself ensure accurate baseline reconstruction. Although the baseline reconstructs the box well, its mean IoUs at $30\,\mathrm{dB}$ are $0.547$, $0.662$, and $0.716$ for the flat plate, trapezoidal prism, and pentagonal prism, respectively, compared with $0.701$, $0.857$, and $0.928$ for the proposed method. The flat plate also exhibits substantial baseline variability, with an IoU standard deviation of $0.157$, compared with $0.028$ for the proposed method. These results are consistent with sensitivity to the estimated normal directions and area coefficients, including their modification during resampling and merging. For the flat plate, the sensitivity of volume overlap to small surface displacements also contributes to the relatively low IoU.

The curved targets exhibit a wide range of baseline performance, with the largest improvement occurring for the capsule. At $30\,\mathrm{dB}$, the baseline achieves mean IoUs of $0.825$, $0.583$, and $0.077$ for the cylinder, capped cone, and capsule, respectively, whereas the proposed method achieves $0.883$, $0.904$, and $0.929$. The cylinder therefore remains reasonably well reconstructed by the baseline, while the capped cone and especially the capsule show much larger gains. This pattern is consistent with the limitations of representing distributed surface-normal information using a sparse set of retained directions. However, the end-to-end comparison changes both EGI recovery and support optimization, so these results establish the advantage of the complete framework without isolating the contribution of either stage.

\begin{table*}[htbp]
\centering
\caption{End-to-end shape-recovery comparison at a signal-to-noise ratio of $30\,\mathrm{dB}$ over 10 reconstruction runs. Values are reported as mean $\pm$ standard deviation. Bold indicates the better mean for each comparison. The final row reports averages of the shape-wise means.}
\label{tab:baseline_comparison_snr30}
\small
\setlength{\tabcolsep}{4pt}
\begin{tabular}{lcccccc}
\toprule
& \multicolumn{2}{c}{IoU $\uparrow$}
& \multicolumn{2}{c}{Hausdorff distance $\downarrow$}
& \multicolumn{2}{c}{Chamfer distance $\downarrow$} \\
\cmidrule(lr){2-3}\cmidrule(lr){4-5}\cmidrule(lr){6-7}
Shape & Baseline & Proposed & Baseline & Proposed & Baseline & Proposed \\
\midrule
Box
& $\mathbf{0.899 \pm 0.026}$ & $0.898 \pm 0.017$
& $0.078 \pm 0.020$ & $\mathbf{0.047 \pm 0.004}$
& $\mathbf{0.042 \pm 0.008}$ & $0.043 \pm 0.006$ \\

Flat plate
& $0.547 \pm 0.157$ & $\mathbf{0.701 \pm 0.028}$
& $0.196 \pm 0.135$ & $\mathbf{0.060 \pm 0.007}$
& $0.072 \pm 0.055$ & $\mathbf{0.043 \pm 0.004}$ \\

Trapezoidal prism
& $0.662 \pm 0.010$ & $\mathbf{0.857 \pm 0.013}$
& $0.154 \pm 0.016$ & $\mathbf{0.046 \pm 0.004}$
& $0.106 \pm 0.004$ & $\mathbf{0.042 \pm 0.004}$ \\

Pentagonal prism
& $0.716 \pm 0.015$ & $\mathbf{0.928 \pm 0.009}$
& $0.183 \pm 0.021$ & $\mathbf{0.043 \pm 0.002}$
& $0.143 \pm 0.009$ & $\mathbf{0.037 \pm 0.003}$ \\

Cylinder
& $0.825 \pm 0.057$ & $\mathbf{0.883 \pm 0.027}$
& $0.087 \pm 0.015$ & $\mathbf{0.042 \pm 0.003}$
& $0.067 \pm 0.022$ & $\mathbf{0.045 \pm 0.009}$ \\

Capped cone
& $0.583 \pm 0.048$ & $\mathbf{0.904 \pm 0.017}$
& $0.364 \pm 0.022$ & $\mathbf{0.067 \pm 0.008}$
& $0.207 \pm 0.027$ & $\mathbf{0.048 \pm 0.007}$ \\

Capsule
& $0.077 \pm 0.017$ & $\mathbf{0.929 \pm 0.008}$
& $0.328 \pm 0.014$ & $\mathbf{0.044 \pm 0.005}$
& $0.530 \pm 0.023$ & $\mathbf{0.032 \pm 0.003}$ \\
\midrule
Mean
& 0.615 & \textbf{0.872}
& 0.198 & \textbf{0.050}
& 0.167 & \textbf{0.041} \\
\bottomrule
\end{tabular}
\end{table*}

\begin{table*}[htbp]
\centering
\caption{IoU comparison across noise levels over 10 reconstruction runs. Values are reported as mean $\pm$ standard deviation. Bold indicates the higher mean for each comparison. The final row reports averages of the shape-wise means. Results at $30\,\mathrm{dB}$ are reported in Table~\ref{tab:baseline_comparison_snr30}.}
\label{tab:baseline_comparison_noise}
\small
\setlength{\tabcolsep}{5pt}
\begin{tabular}{lcccccc}
\toprule
& \multicolumn{2}{c}{Noiseless}
& \multicolumn{2}{c}{$40\,\mathrm{dB}$}
& \multicolumn{2}{c}{$20\,\mathrm{dB}$} \\
\cmidrule(lr){2-3}\cmidrule(lr){4-5}\cmidrule(lr){6-7}
Shape & Baseline & Proposed & Baseline & Proposed & Baseline & Proposed \\
\midrule
Box
& $\mathbf{0.915 \pm 0.027}$ & $0.895 \pm 0.019$
& $\mathbf{0.913 \pm 0.019}$ & $0.896 \pm 0.018$
& $\mathbf{0.894 \pm 0.025}$ & $0.893 \pm 0.014$ \\

Flat plate
& $0.563 \pm 0.070$ & $\mathbf{0.697 \pm 0.022}$
& $0.556 \pm 0.141$ & $\mathbf{0.699 \pm 0.024}$
& $0.594 \pm 0.120$ & $\mathbf{0.708 \pm 0.030}$ \\

Trapezoidal prism
& $0.658 \pm 0.007$ & $\mathbf{0.859 \pm 0.017}$
& $0.659 \pm 0.005$ & $\mathbf{0.860 \pm 0.015}$
& $0.658 \pm 0.009$ & $\mathbf{0.860 \pm 0.012}$ \\

Pentagonal prism
& $0.723 \pm 0.018$ & $\mathbf{0.929 \pm 0.006}$
& $0.721 \pm 0.021$ & $\mathbf{0.928 \pm 0.007}$
& $0.713 \pm 0.017$ & $\mathbf{0.924 \pm 0.014}$ \\

Cylinder
& $0.807 \pm 0.068$ & $\mathbf{0.897 \pm 0.006}$
& $0.814 \pm 0.079$ & $\mathbf{0.893 \pm 0.008}$
& $0.791 \pm 0.045$ & $\mathbf{0.829 \pm 0.163}$ \\

Capped cone
& $0.563 \pm 0.049$ & $\mathbf{0.950 \pm 0.013}$
& $0.563 \pm 0.041$ & $\mathbf{0.935 \pm 0.010}$
& $0.550 \pm 0.048$ & $\mathbf{0.850 \pm 0.022}$ \\

Capsule
& $0.072 \pm 0.014$ & $\mathbf{0.938 \pm 0.014}$
& $0.083 \pm 0.017$ & $\mathbf{0.943 \pm 0.013}$
& $0.092 \pm 0.037$ & $\mathbf{0.914 \pm 0.011}$ \\
\midrule
Mean
& 0.614 & \textbf{0.881}
& 0.616 & \textbf{0.879}
& 0.613 & \textbf{0.854} \\
\bottomrule
\end{tabular}
\end{table*}

\FloatBarrier
\subsection{Controlled Stage Comparison}
\label{sec:controlled_stage_comparison}

The contributions of EGI recovery and support optimization are examined by comparing all four combinations of the baseline and proposed stages. Table~\ref{tab:stage_configurations} defines configurations M1--M4, with M1 and M4 corresponding to the complete baseline and proposed pipelines, respectively. The EGI recovery stage uses either NNLS (Section~\ref{sec:baseline-egi}) or the proposed {denoiser-regularized} method (Section~\ref{sec:plug-and-play-image-recovery}), while support optimization uses either the constrained trust-region method (Section~\ref{sec:baseline-support}) or the scale-invariant method (Section~\ref{sec:proposed-shape-recovery}). Each configuration was evaluated at $30\,\mathrm{dB}$ over 10 paired reconstruction runs per shape. For each EGI recovery method, the same recovered EGI was supplied to both support solvers, allowing their performance to be compared without variation in the input estimate. 

The proposed EGI recovery method produces a substantial aggregate improvement even when the baseline support solver is retained. As shown in Table~\ref{tab:stage_comparison}, replacing NNLS with {denoiser-regularized} recovery increases the mean IoU from $0.615$ for M1 to $0.804$ for M3. The largest improvement occurs for the capsule, whose mean IoU increases from $0.077$ to $0.928$, nearly matching the complete proposed pipeline. However, this improvement is not uniform across geometries. M3 yields lower mean IoU than M1 for the box and flat plate. Thus, improved aggregate performance from the proposed EGI recovery does not guarantee better reconstruction for every shape under the baseline support solver. 

The benefit of scale-invariant support optimization depends strongly on the EGI recovery method. With NNLS-derived EGIs, M1 and M2 have nearly identical mean IoUs of $0.615$ and $0.616$, and their shape-wise means differ by at most $0.002$. With the proposed EGIs, replacing the constrained trust-region solver with the scale-invariant solver increases the mean IoU from $0.804$ for M3 to $0.872$ for M4. The largest gains occur for the flat plate and box, with increases of $0.200$ and $0.109$, respectively, whereas the capsule changes by only $0.001$. The support solver therefore provides a substantial additional benefit for some proposed EGIs while having little effect on others.

The controlled comparison demonstrates that EGI recovery and support optimization interact in determining reconstruction accuracy. EGI recovery accounts for the larger aggregate change in this experiment, but the complete proposed pipeline also benefits from the numerical strategy used to reconstruct its recovered EGIs. Because the effect of changing one stage depends on the other stage, the overall improvement cannot be assigned a unique percentage contribution from each component.

\begin{table}[htbp]
\centering
\caption{Method configurations used in the controlled stage comparison. M1 and M4 correspond to the baseline and proposed pipelines, respectively; M2 and M3 exchange one recovery stage.}
\label{tab:stage_configurations}
\small
\setlength{\tabcolsep}{8pt}
\begin{tabular}{lll}
\toprule
Name & EGI recovery & Support optimization \\
\midrule
M1 (Baseline) & NNLS & Constrained trust-region method \\
M2 & NNLS & Scale-invariant \\
M3 & {Denoiser-regularized} & Constrained trust-region method \\
M4 (Proposed) & {Denoiser-regularized} & Scale-invariant \\
\bottomrule
\end{tabular}
\end{table}

\begin{table*}[htbp]
\centering
\caption{IoU for the four configurations in Table~\ref{tab:stage_configurations} at $30\,\mathrm{dB}$ over 10 paired reconstruction runs per shape. Values are mean $\pm$ standard deviation; the final row averages the shape-wise means. Bold indicates the highest mean at the reported precision.}
\label{tab:stage_comparison}
\small
\setlength{\tabcolsep}{8pt}
\begin{tabular}{lcccc}
\toprule
Shape & M1 (Baseline) & M2 & M3 & M4 (Proposed) \\
\midrule
Box & $\mathbf{0.899 \pm 0.026}$ & $\mathbf{0.899 \pm 0.026}$ & $0.789 \pm 0.154$ & $0.898 \pm 0.017$ \\
Flat plate & $0.547 \pm 0.157$ & $0.546 \pm 0.157$ & $0.501 \pm 0.122$ & $\mathbf{0.701 \pm 0.028}$ \\
Trapezoidal prism & $0.662 \pm 0.010$ & $0.662 \pm 0.010$ & $0.805 \pm 0.043$ & $\mathbf{0.857 \pm 0.013}$ \\
Pentagonal prism & $0.716 \pm 0.015$ & $0.715 \pm 0.014$ & $0.912 \pm 0.028$ & $\mathbf{0.928 \pm 0.009}$ \\
Cylinder & $0.825 \pm 0.057$ & $0.827 \pm 0.056$ & $0.828 \pm 0.020$ & $\mathbf{0.883 \pm 0.027}$ \\
Capped cone & $0.583 \pm 0.048$ & $0.583 \pm 0.048$ & $0.867 \pm 0.025$ & $\mathbf{0.904 \pm 0.017}$ \\
Capsule & $0.077 \pm 0.017$ & $0.077 \pm 0.017$ & $0.928 \pm 0.015$ & $\mathbf{0.929 \pm 0.008}$ \\
\midrule
Mean & 0.615 & 0.616 & 0.804 & \textbf{0.872} \\
\bottomrule
\end{tabular}
\end{table*}

\FloatBarrier
\section{Conclusions}
\label{sec:conclusions}

This study developed a {denoiser-regularized} framework for reconstructing convex shapes from unresolved light-curve measurements. Representing the EGI as an area image enables the use of PnP-ADMM with a domain-adapted DRUNet denoiser to estimate distributed surface-normal information. The recovered EGI is converted into a convex polytope through scale-invariant support optimization with analytical gradients and explicit volume normalization. This support formulation preserves the geometric recovery principle of the baseline while providing an alternative numerical solution strategy. Random-frame recovery varies the alignment between the EGI and the equirectangular grid, with the resulting variability included in evaluation over repeated reconstruction runs. 

The numerical experiments demonstrate consistent aggregate reconstruction performance across the tested geometries and noise conditions. Over seven shapes and 10 runs per shape and condition, mean IoU values were $0.881$, $0.879$, $0.872$, and $0.854$ for noiseless measurements and SNRs of $40$, $30$, and $20\,\mathrm{dB}$, respectively. Relative to the baseline, the proposed framework increased mean IoU at $30\,\mathrm{dB}$ from $0.615$ to $0.872$, while reducing mean Hausdorff distance from $0.198$ to $0.050$ and mean Chamfer distance from $0.167$ to $0.041$. All three metrics improved on six of the seven shapes. The box remained competitive under the baseline, whereas the capped cone and capsule exhibited particularly large gains in IoU.

The controlled stage comparison shows that the improvement depends on both EGI recovery and its subsequent reconstruction. Substituting the proposed EGI recovery while retaining the baseline support solver increased mean IoU from $0.615$ to $0.804$ at $30\,\mathrm{dB}$. Using the proposed support solver with those EGIs further increased mean IoU to $0.872$. In contrast, changing the support solver for NNLS-derived EGIs produced little change in mean IoU. These results identify EGI recovery as the source of the larger aggregate improvement and demonstrate an additional, input-dependent benefit from the proposed support optimization strategy.

The remaining limitations concern coordinate sensitivity and the assumptions of the forward model. Although random-frame recovery improved the preliminary flat-plate results, it does not guarantee favorable grid alignment, and the cylinder exhibited intermittent orientation-sensitive failures at $20\,\mathrm{dB}$. A preliminary systematic alignment strategy reduced variability in that case but did not consistently improve reconstruction accuracy across shapes. The present experiments also assume convex geometry, known attitude and observation geometry, and known homogeneous surface-reflectance properties. Future work will investigate representations and frame-selection strategies that reduce coordinate-dependent artifacts, account for attitude and reflectance uncertainty, and extend the framework to more general spacecraft geometries and observational light curves.

\section*{Acknowledgment}
This work was supported by the Air Force Office of Scientific Research (AFOSR) under Grant No. FA9550-18-1-0154. The authors thank Liam Robinson at Purdue University for providing the source code used to implement the baseline method.

\section*{Declaration of generative AI and AI-assisted technologies in the manuscript preparation process}
During the preparation of this work, the authors used Claude and ChatGPT to check for consistency in notation, reconcile bibliography entries, and improve the organization and grammar of the draft. The authors reviewed and edited the output as needed and take full responsibility for the content of the published article.


\end{document}